\documentclass{cimento}
\usepackage{amsmath}
\usepackage{amssymb}
\usepackage{graphicx}
\numberwithin{equation}{section}

\title{The High Temperature Superconductivity in Cuprates}
\author{Paolo Cea\from{inst1}\from{inst2}\thanks{Paolo.Cea@ba.infn.it}}
\instlist{\inst{inst1} Dipartimento di Fisica dell'Universit\`a di Bari, I-70126 Bari, Italy 

            \inst{inst2} INFN - Sezione di Bari, I-70126 Bari, Italy}

\PACSes{
\PACSit{74.20.-z, 74.72.-h, 74.72.Gh, 74.72.Ek}{}
}

\begin{document}

\maketitle

\begin{abstract}
We discuss the high-temperature superconductivity in copper oxide ceramics. We propose an effective Hamiltonian to describe the
dynamics of electrons or holes injected into the copper oxide layers. We show that our approach is able to account for both
the pseudogap and the superconductivity gap. For the hole-doped cuprates we discuss in details the underdoped, optimal doped, and 
overdoped regions of the phase diagram. In the underdoped region we determine the doping dependence of the upper critical
magnetic field, the vortex region, and the discrete states bounded to the core of isolated vortices. We explain the origin of the Fermi
arcs and Fermi pockets. Moreover, we discuss the recently reported peculiar dependence of the specific heat on the applied magnetic field.
We determine the critical doping where the pseudogap vanishes. We find that in the overdoped region the superconducting transition is 
described by the conventional d-wave BCS theory. We discuss the optimal doping region and the crossover between the underdoped region
and the overdoped region. We also discuss briefly the electron-doped cuprate superconductors.
\end{abstract}

\section{Introduction}
\label{1}

\begin{figure}[t]
\begin{center} 
\includegraphics[width=0.8\textwidth,clip]{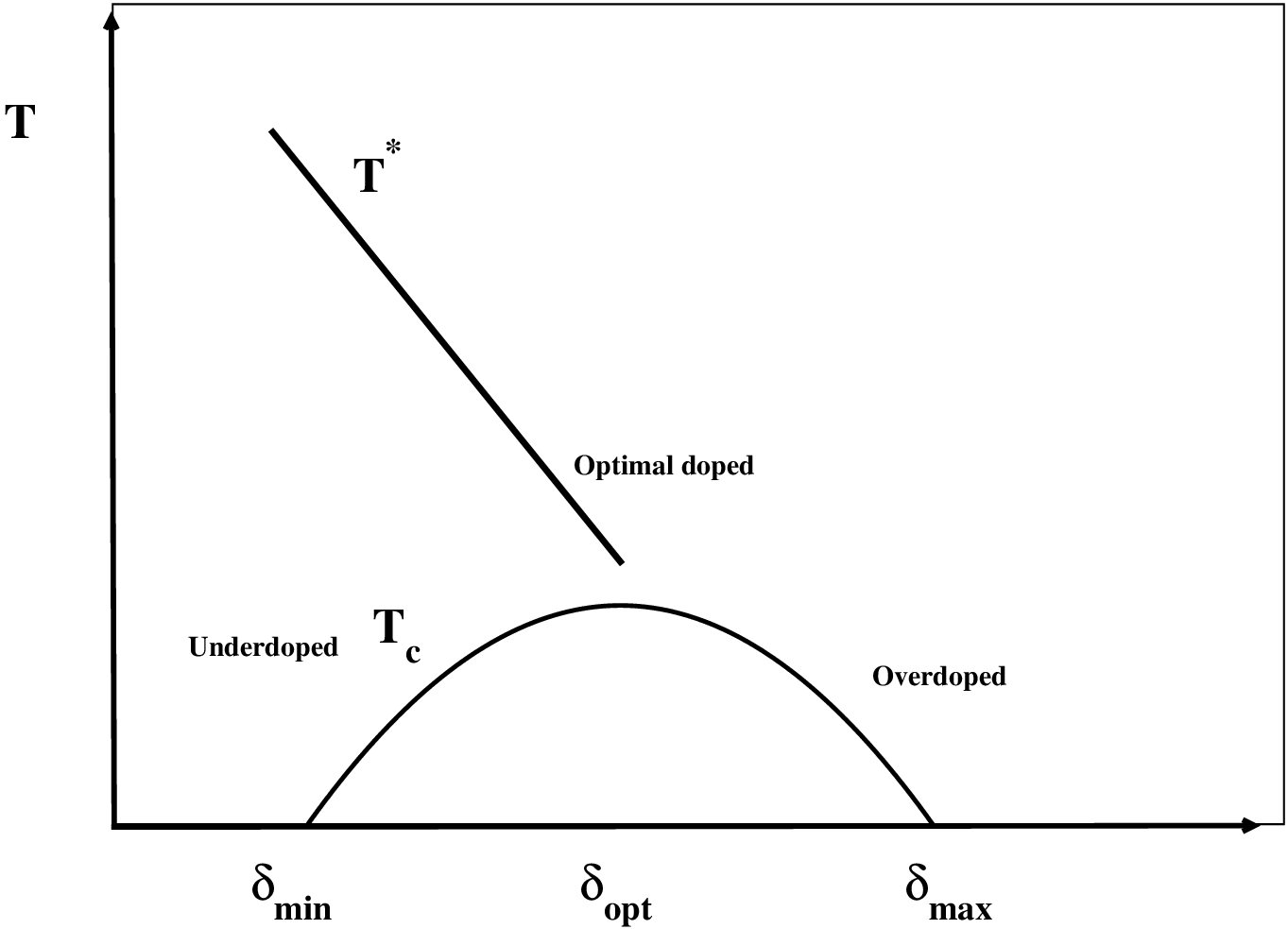}
\caption{\label{fig-1}  
Schematic phase diagram of hole-doped cuprate high-temperature superconductors. Carrier concentrations
$\delta$ near to the maximum critical temperature $T_c(\delta_{opt})$ locate the optimal doped region. Regions with lower and larger hole concentration are called
underdoped and overdoped, respectively. The temperature $T^*$ is the pseudogap temperature. }
\end{center}
\end{figure}

Since the discovery of superconductivity in the copper oxide ceramics (cuprates)~\cite{Bednorz:1986}, both theoretical and experimental
efforts have been made to investigate the physical mechanism responsible for superconductivity with an  unprecedented high transition temperature
$T_c$ (for recent reviews see Refs.~\cite{Damascelli:2003,Deutscher:2005,Besov:2005,Lee:2006,Fischer:2007,Lee:2008,Hufner:2008,Sebastian:2012}).
Despite this, an understanding of the superconducting pairing of the high-temperature superconductors is still lacking. \\
Very soon after the discovery of the high temperature superconductors, it was realized~\cite{Anderson:1987} that superconductivity were
intimately related to the square planar $CuO_2$ lattice. Moreover, the physics of the $CuO_2$  planes were well described by the nearly
half-filled Hubbard model with moderately on-site Coulomb repulsion. \\
The microscopic model for the description of electrons in the    $CuO_2$  layers is believed to be the effective single-band Hubbard 
model:~\footnote{For a lucid discussion see Ref.~\cite{Anderson:1997}.}
\begin{equation}
\label{1.1}
\hat{H} \; = \; -t \; \sum_{<i,j>,\sigma}   \left [  \hat{c}^{\dagger}_{i,\sigma}  \, \hat{c}_{j,\sigma}  \,  + \,   \hat{c}^{\dagger}_{j,\sigma}  \, \hat{c}_{i,\sigma} \right] \, + \,
U   \,  \sum_{i}   \hat{n}_{i,\uparrow}  \, \hat{n}_{i,\downarrow}  \; \; , \; \;   \hat{n}_{i,\sigma} \, = \,  \hat{c}^{\dagger}_{i,\sigma}  \, \hat{c}_{i,\sigma}  \; ,
\end{equation}
where $ \hat{c}^{\dagger}_{i,\sigma} $ and   $ \hat{c}_{i,\sigma} $ are creation and annihilation operators for electrons with spin $\sigma$, $U$ is the onsite Coulomb
repulsion for electrons of opposite spin at the same atomic orbital, and $t$ is the hopping parameter.  As is well known~\cite{Anderson:1959}, the superexchange mechanism 
yields a Heisenberg antiferromagnetic exchange interaction with $J \, = \, \frac{4 t^2}{U}$ between spins on the $Cu$ atoms.  Since $ U \, \gg t$, at half-filling the onsite Coulomb repulsion gives an insulator where the electron spins are antiferromagnetically ordered. \\
The phase diagram resulting from tuning the doping of  $CuO_2$  planes with holes is shown schematically in Fig.~\ref{fig-1}. At half-filling, corresponding to $\delta = 0$,
we have a Mott insulator with long range antiferromagnetic order which disappears for temperatures above the critical N\'eel temperature $T_N$ (not shown in Fig.~\ref{fig-1}).
By adding holes into the  $CuO_2$  planes ($\delta > 0$) the long range antiferromagnetic order is rapidly destroyed. Superconductivity occurs beyond a minimal
hole content $\delta_{min} \simeq 0.05 - 0.06$ where there is no long-range antiferromagnetic order. Nevertheless, signs of magnetism persist~\cite{Bonn:2006}
in the hole-doped materials with $\delta > \delta_{min}$ as local order on a  short length scale $\xi_{AF} \gtrsim  10^{-6} cm$, as observed in muon spin 
relaxation~\cite{Niedermayer:1998} and neutron scattering~\cite{Cheong:1991,Mason:1992,Lavrov:2009} measurements. \\
The superconducting critical temperature $T_c$ rises with increasing hole doping through the underdoped region ($\delta < \delta_{opt}$), reaches a peak at
optimal doping $\delta \simeq \delta_{opt}$, falls away at higher doping $\delta > \delta_{opt}$ (overdoped region), and then vanishes at the maximal
doping  $\delta_{max}  \simeq 0.3$. This peculiar doping dependence of the critical temperature $T_c$ is a generic feature of the hole-doped copper oxides.
Another generic feature of the high-temperature cuprate superconductors is the presence of a pseudogap $k_B T^*$ ($k_B$ being the Boltzmann's constant) in
the underdoped region  $\delta  \lesssim \delta_{opt}$. The pseudogap was first detected in the temperature dependence of the spin-lattice relaxation, Knight shift,
and magnetic susceptibility studies~\footnote{ See Ref.~\cite{Hufner:2008} and references therein.}. The pseudogap temperature $T^*$ is much larger
than the superconducting critical temperature $T_c$ in the underdoped region. Moreover, $T^*(\delta)$ decreases upon increasing the doping and seems to merge
with $T_c(\delta)$ for $\delta \simeq \delta_{opt}$. In the overdoped region, however, the coexistence of the superconducting gap with the pseudogap is still
subject to an intense debate. \\
The challenge is to find a theory that is rooted in the microscopic physics of the copper oxides which is able to account for the physics of the pseudogap and
the superconductivity. One of the problem in constructing a satisfactory microscopic theory of the high-temperature superconductivity is to isolate that part
of the interaction which is responsible for both the pseudogap and the superconductivity in cuprates. The purpose of this paper is to present an approach
which, according to the spirit of the conventional superconductivity theory of Bardeen, Cooper, and Schrieffer~\cite{Bardeen:1957} (BCS), allows to isolate
an effective Hamiltonian governing the dynamics of the electrons or holes injected into the undoped copper-oxide planes. We shall assume that the
superconductivity originates into the $CuO_2$  planes, and thereby we shall neglect the motion along the direction perpendicular to the planes. We further
assume that electron dynamics in the $CuO_2$  planes is described by the single-band effective Hubbard model with Hamiltonian given by Eq.~(\ref{1.1}).
Since the majority of high-temperature cuprate superconductors are hole-doped, we shall focus mainly on hole-doped cuprates. Nevertheless, in Section~\ref{6}
we briefly discuss also the electron-doped cuprates (for a recent review, see Ref.~\cite{Armitage:2010}). \\
\indent
The plan of the paper is as follows. In Sect.~\ref{2} we set up the effective interaction between two holes in an antiferromagnetic background and
we discuss the main assumptions of the paper.  Section~\ref{3} is devoted to a general discussion of the physical structure of the superconducting
ground state. In Sect.~\ref{3-1} we discuss the real-space d-wave bound states of two holes;  in  Sect.~\ref{3-2} we address the problem 
of the superconductivity and pseudogap in the underdoped region. In Sect.~\ref{3-3} we discuss the d-wave BCS ground state and argue that it is relevant
in the overdoped region. In Sect.~\ref{3-4} we present our understanding of the phase diagram of hole-doped cuprates. 
The physics of the underdoped region is presented in Sect.~\ref{4} where we discuss the real-space bound states in an external magnetic field (Sect.~\ref{4-1}),
the upper critical magnetic field  (Sect.~\ref{4-2}), the vortex region (Sect.~\ref{4-3}), the physics of Fermi arcs and quantum oscillations  (Sect.~\ref{4-4}).
Section~\ref{5} is devoted to the discussion of the physics in the overdoped and optimal doped regions together with the crossover between the underdoped region
and the overdoped region. In Sect.~\ref{6} we briefly address the problem of superconductivity and pseudogap in the electron-doped cuprates. 
Finally, in Section~\ref{7} we summarize the main achievements of the paper and we draw our conclusions.
\section{The effective Hamiltonian}
\label{2}
\begin{figure}[t]
\begin{center} 
\includegraphics[width=0.8\textwidth,clip]{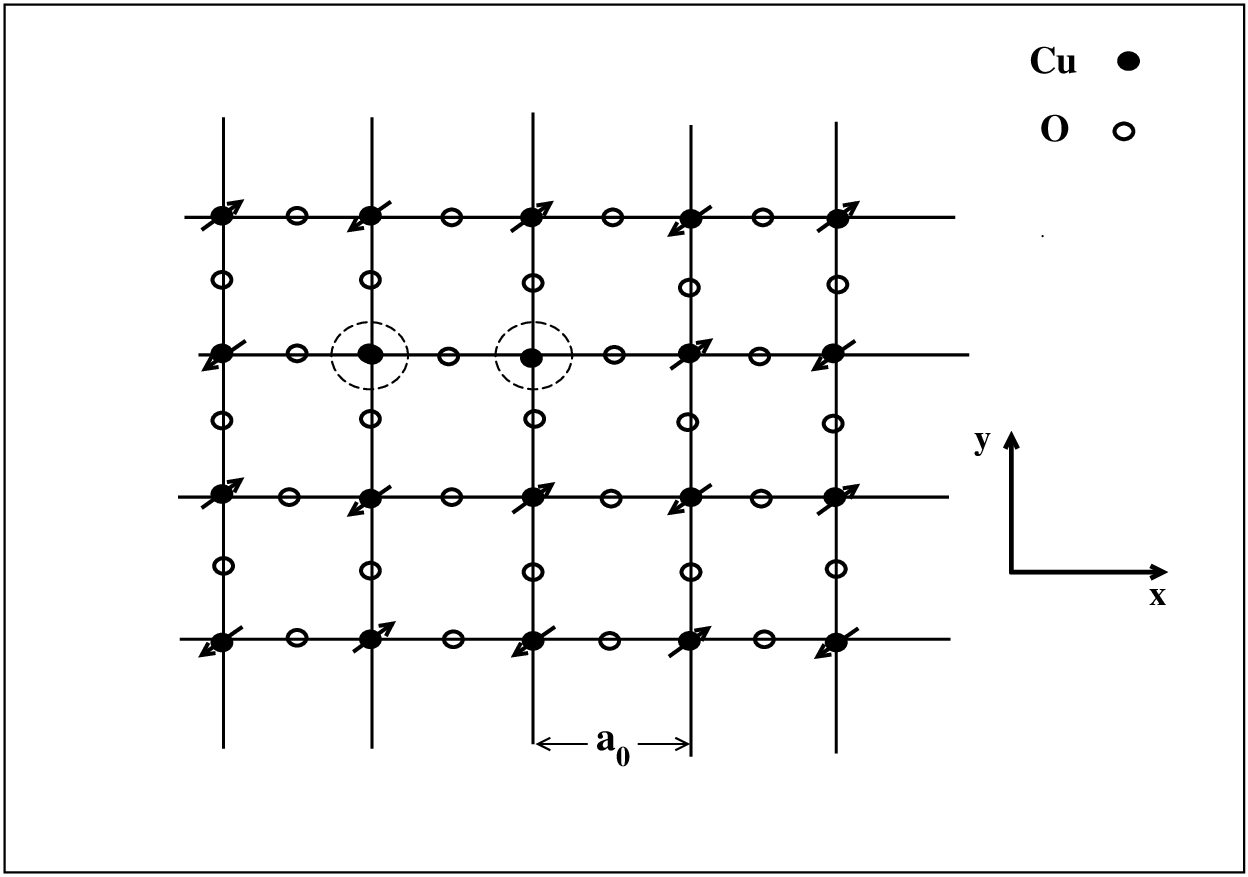}
\caption{\label{fig-2}  The idealized square copper plane with lattice spacing $a_0$. The dashed ellipses represent two holes in an antiferromagnetic background. }
\end{center}
\end{figure}

In this Section we set up our effective Hamiltonian for the dynamics of holes in the nearly half-filled single-band Hubbard model. As discussed in the Introduction, we are
assuming that the superconductivity in cuprates is related to the square planar $CuO_2$ lattice. The single-band Hamiltonian is given by Eq.~(\ref{1.1}), where $<i,j>$ 
denotes a pair of nearest-neighbor sites of a square lattice with lattice constant $a_0$ (see Fig.~\ref{fig-2}). Let $M$ be the number of sites of the copper lattice and $N$
the number of electrons. In the half filled lattice $N=M$, i.e. there is exactly one electron per site. In the underfilled case $N < M$ there are holes with doping fraction
(number of holes per site):
\begin{equation}
\label{2.1}
\delta \; = \; 1 \; - \; \frac{N}{M}  \; .
\end{equation}
In the subspace of the state space of avoided double occupancies of $Cu$ sites, the one-band Hubbard model may be canonically transformed into the so-called
$t-J$ Hamiltonian:
\begin{equation}
\label{2.2}
\hat{H} \; = \; -t \; \sum_{<i,j>,\sigma}   \left [  \hat{\cal{C}}^{\dagger}_{i,\sigma}  \, \hat{\cal{C}}_{j,\sigma}  \,  + \,   \hat{\cal{C}}^{\dagger}_{j,\sigma}  \, \hat{\cal{C}}_{i,\sigma} \right] \, + \,
J   \,  \sum_{<i,j>}   \left [  {\vec{S}}_i   \cdot  { \vec{S}}_j  \; - \; \frac{1}{4} \; \hat{n}_{i}  \, \hat{n}_{j}  \right] \; \; , 
\end{equation}
where
\begin{equation}
\label{2.3}
 \hat{\cal{C}}_{i,\sigma} \; = \;    \hat{c}_{i,\sigma} \, ( 1 \, - \, \hat{n}_{i,-\sigma}) \; \; , \; \; J \; = \; \frac{4 t^2}{U} \; ,
\end{equation}
and $\vec{S}_i$ is the spin-$\frac{1}{2}$ operator at site $i$.~\footnote{ See, for instance,  Ref.~\cite{Balachandran:1990} and references therein.}
Since at half-filling $ \hat{n}_{i,-\sigma} = 1$, the $t-J$ Hamiltonian Eq.~(\ref{2.2}) reduces to the square lattice Heisenberg Hamiltonian:
\begin{equation}
\label{2.4}
\hat{H} \; = \;  J   \,  \sum_{<i,j>}     {\vec{S}}_i   \cdot  { \vec{S}}_j  \;  \; \; . 
\end{equation}
The ground state of the latter Hamiltonian is known to be the antiferromagnetic N\'eel state. When the system is doped with holes away from half filling
($\delta > 0$), the dynamics is no longer described by the Heisenberg Hamiltonian  Eq.~(\ref{2.4}), but one must use the $t-J$ Hamiltonian  Eq.~(\ref{2.2})
with the supplementary constraint of no double occupied lattice sites. It is easy to see that the motion of dilute holes in the antiferromagnetic N\'eel background
is strongly frustrate. Indeed, when a hole hops from one site to its nearest neighbor, a spin is also moved onto one lattice site with the wrong spin orientation
with respect to the antiferromagnetic background. On the other hand, a pair of holes with nearest-neighbor distance (see Fig.~\ref{fig-2}) can move almost freely
without destroying the antiferromagnetic background. The effective Hamiltonian for the propagation of the holes in the antiferromagnetic background
can be written as~\cite{Huang:1987,Hirsch:1987}:
\begin{equation}
\label{2.5}
\hat{H}_0  \; = \; - \frac{t^2}{U}  \; \sum_{\vec{r}}  \sum_{i,j}   \hat{\psi}^{\dagger}_{h}(\vec{r} + \vec{i} a_0 + \vec{j} a_0) \hat{\psi}_{h}(\vec{r})  \; ,
\end{equation}
where $ \hat{\psi}^{\dagger}_{h}(\vec{r})$, $ \hat{\psi}_{h}(\vec{r})$ are creation and annihilation operators for holes at the lattice site $\vec{r}$ and the sum over
the direction vectors $ \vec{i}$ and $\vec{j}$ is restricted to next-nearest neighbor lattice sites. Note that in  the effective Hamiltonian $\hat{H}_0$  we do not display
the spin of the holes since the antiferromagnetic background forces the two  holes to have antiparallel spins. In other words, the motion of a hole must not disturb the 
antiferromagnetic background. Moreover, in this antiferromagnetic background approximation, it turns out that there is an effective attractive two-body potential between
nearest-neighbor holes~\cite{Huang:1987}. However, it has been pointed out~\cite{Trugman:1988} that the motion of a pair of holes, which naively should be mobile, is
frustrated due to fermionic nature of the background spins. Indeed, for states with two holes as nearest neighbors in an antiferromagnetic background it is easy to 
see~\cite{Carlson:2002} that there are  configurations that differ by the exchange of two electrons. The destructive interference of these configurations leads
to holes with very large effective mass. This problem can be avoided if we consider pairs of hole with distance much greater than the lattice spacing. Accordingly,
we shall suppose that two holes with distance $a_0 \ll r_0 \ll \xi_{AF}$, where $\xi_{AF}$ is the antiferromagnetic local order length scale,  are subject to an effective attractive
two-body potential. Our proposal is quite similar to the spin-bag theory~\cite{Schrieffer:1988} where the pairing is due to a local reduction of the antiferromagnetic order (bag)  shared by two holes. \\
Thus, we are lead to consider the following reduced interaction Hamiltonian:
\begin{equation}
\label{2.6}
\hat{H}_{int} \;  =  \; \frac{1}{2}   \int d \vec{r}_1\,  d \vec{r}_2  \; \hat{\psi}^{\dagger}_{h,\uparrow}(\vec{r}_1)   \,   \hat{\psi}^{\dagger}_{h,\downarrow}(\vec{r}_2) 
\, V(\vec{r}_1 - \vec{r}_2)  \, \hat{\psi}_{h,\downarrow}(\vec{r}_2)  \,  \hat{\psi}_{h,\uparrow}(\vec{r}_1) \; .
\end{equation}
The simplest choice for the two-body potential $V(\vec{r_1} - \vec{r_2})$ is:
\begin{equation}
\label{2.7} 
V(\vec{r_1} - \vec{r_2}) \;  = \; 
 \left\{ \begin{array}{ll}
 \; \; \infty \; \; \; \; &  \vec{r_1} \, = \, \vec{r_2}
  \\
  - \, V_0  \; \; \; \; &  |\vec{r_1} - \vec{r_2}| \, \leq \, r_0(\delta)
  \\
  \; \; \;  \;  0    \; \; \; \;  & otherwise
\end{array}
    \right.
\end{equation}
where the hard-core at the origin is imposed to avoid configurations where two holes are on the same site. Moreover, since we are considering the limit of low
carriers, $\delta \ll 1$, the hard-core potential does not allow two holes to come too close, thereby implementing the no-double occupation constraint. As concern 
the range of our potential $ r_0(\delta)$, it is natural to choice  $ r_0(\delta)$ of the order of the coherence length $\xi_0 \simeq 6 a_0$. So that we assume:
\begin{equation}
\label{2.8} 
 r_0(\delta) \; = \;  6 \;  a_0 \; \left ( 1 \; - \; \frac{\delta}{\delta_c} \right )^{\frac{1}{2}} \; \; \; , \; \; \delta_c \; \simeq \; 0.35 \; .
 \end{equation}
The peculiar dependence on the doping fraction $\delta$ arises as follows. The attractive two-body potential originates from the interaction of two holes in
the antiferromagnetic background which extends over a distance of order $\xi_{AF}$. Since the injected holes tend to destruct the antiferromagnetic order,
the area of the antiferromagnetic islands should decrease with increasing $\delta$. In addition, as discussed in the Introduction (see Fig.~\ref{fig-1}), the
superconducting instability disappears at the maximal doping $\delta_{max} \simeq 0.30$. The simplest parameterization which takes care of these effects leads to our
Eq.~(\ref{2.8}). Regarding the parameter $V_0$ in Eq.~(\ref{2.7}), we have fixed this parameter such that there is at least one real space d-wave bound state.
To this end it suffices to assume:
\begin{equation}
\label{2.9} 
 V_0 \; \simeq  \;  2 \;  J \; = \; \frac{8 t^2}{U} \; .
 \end{equation}
In the following we will use the numerical values:
\begin{equation}
\label{2.10} 
 a_0 \; \simeq \; 4.0 \, 10^{-8} \, cm \; \; \; , \; \; \;  t \; \simeq \; 0.11 \; ev \; \; , \; \;  U \; \simeq  \; 1.1 \; ev \; \; ,  
 \end{equation}
that are appropriate for a typical cuprate. Note that the resulting N\'eel temperature:
\begin{equation}
\label{2.11} 
 T_N \; \simeq \;  \frac{J}{k_B}  \;  \sim  \, 10^{3} \, K  
 \end{equation}
is of the order of the observed N\'eel temperatures in cuprates. \\
Before concluding this Section, we would like to rewrite the Hamiltonian $\hat{H}_0$  Eq.~(\ref{2.5}) in the limit of low-lying excitations. Observing that
$\hat{H}_0$ can be diagonalized by writing:
\begin{equation}
\label{2.12} 
 \hat{\psi}_{h}(\vec{r})  \; = \;  \frac{1}{\sqrt{M}}  \; \sum_{\vec{k}}  \exp{(i \vec{k} \cdot \vec{r})}    \; \hat{\psi}_{h}(\vec{k}) \; , 
 \end{equation}
so that:
\begin{equation}
\label{2.13} 
 \hat{\psi}_{h}(\vec{k})  \; = \;  \frac{1}{\sqrt{M}}  \; \sum_{\vec{r}}  \exp{(- i \vec{k} \cdot \vec{r})}    \; \hat{\psi}_{h}(\vec{r}) \; .
 \end{equation}
We get:
\begin{equation}
\label{2.14} 
\hat{H}_0  \; = \;    \; \sum_{\vec{k}}  \varepsilon_{\vec{k}}   \;  \;  \hat{\psi}^{\dagger}_{h}(\vec{k})  \,   \hat{\psi}_{h}(\vec{k}) \; ,
 \end{equation}
where:
\begin{equation}
\label{2.15} 
\varepsilon_{\vec{k}}  \; = \;  -  \; \frac{2 t^2}{U}  \; \left [ \; \cos{(2 k_x a_0}) \; +    \cos{(2 k_y a_0}) \;  \right ] \; . 
 \end{equation}
We now take the small-k limit:
\begin{equation}
\label{2.16} 
\hat{H}_0  \; = \;    \; \sum_{\vec{k}}  \frac{\hbar^2 \, \vec{k}^2}{2 \, m^*_h}   \;  \;  \hat{\psi}^{\dagger}_{h}(\vec{k})  \,   \hat{\psi}_{h}(\vec{k}) 
 \end{equation}
with:
\begin{equation}
\label{2.17} 
m^*_h \; = \;  \frac{\hbar^2}{8  \, \frac{t^2}{U} a_0^2} \; . 
\end{equation}
So that  we may write for the effective Hamiltonian:
\begin{equation}
\label{2.18} 
\hat{H}_0  \; = \;    \; \sum_{\sigma=\uparrow, \downarrow }    \int d \vec{r} \;   \hat{\psi}^{\dagger}_{h,\sigma}(\vec{r}) \, \left ( \frac{- \hbar^2 \nabla^2}{2 \, m^*_h}
\, \right )  \hat{\psi}_{h,\sigma}(\vec{r}) \; ,
\end{equation}
where we make explicit the spin degrees of freedom. Using the numerical values of the parameters we obtain:
\begin{equation}
\label{2.19} 
m^*_h \; \simeq \;  5.41 \; m_e \; , 
\end{equation}
where $m_e$ is the electron mass. To summarize, our effective Hamiltonian for low-lying excitations may be described by:
\begin{equation}
\label{2.20} 
\hat{H}_{eff}  \; = \;   \hat{H}_0 \; \; + \; \; \hat{H}_{int} \; ,
\end{equation}
with $\hat{H}_0$ and $\hat{H}_{int}$ given by Eqs.~(\ref{2.18}) and (\ref{2.6}) respectively.  It is worthwhile to stress that the arguments developed in this Section 
cannot be considered as a truly microscopic derivation of the effective Hamiltonian  Eq.~(\ref{2.20}). 
\section{The superconducting ground state}
\label{3}
For later convenience it is useful to think about the wave function of the superconducting ground state within the BCS theory.
Let $\varphi(\vec{r}_1-\vec{r}_2)$ be the wavefunction of a Cooper pair~\cite{Cooper:1956}, thus the superconducting ground state may be expressed
as an antisymmetrized product of identical pair functions: 
\begin{equation}
\label{3.1} 
\Omega(\vec{r}_1, ..., \vec{r}_N)  \; = \;  \sum_P \; (-1)^P \; P \left \{ \varphi(\vec{r}_1-\vec{r}_2)  \cdot \, \cdot \, \cdot   \varphi(\vec{r}_{N-1}-\vec{r}_N) \right \} \; ,
\end{equation}
in which the sum is over all permutations P of the N particles. The expression Eq.~(\ref{3.1}) may be used to point out the analogies and differences between the
superconducting ground state and that of a Bose-Einstein condensation of pairs~\cite{Schafroth:1955,Blatt:1955}. Indeed, a Bose-Einstein condensate of pairs
would also be given as a product of identical pair wavefunctions. The process of antisymmetrizing would not change the character of the state very much if the 
size of the pair wavefunction, $\xi_0$, were small compared with the average spacing between pairs. In this case, wavefunctions in the sum differing by the exchange
of single members of two or more pairs would not overlap very much, and the system would behave qualitatively like a condensed Bose-Einstein gas. However,
antisymmetrization makes a major difference if the size of the pair wavefunction is large compared with the spacing between pairs. In this case, it is convenient to
rewrite the superconducting ground-state wavefunction Eq.~(\ref{3.1})
 in the second quantization formalism. To this end, we introduce the creation operator of a pair at rest: 
\begin{equation}
\label{3.2} 
{ \it{b}^{\dagger}} \; = \; \int d \vec{r}_1   d \vec{r}_2 \;  \varphi(\vec{r}_1-\vec{r}_2) \;    \hat{\psi}^{\dagger}_{\uparrow}(\vec{r_1})   \,   \hat{\psi}^{\dagger}_{\downarrow}(\vec{r_2})  \; .
\end{equation}
Then, the wavefunction Eq.~(\ref{3.1}) can be written as:
\begin{equation}
\label{3.3} 
 | \Omega > _{N}  \; = \;  \frac{  ({ \it{b}^{\dagger}})^{N'}  }{N' !} \; | 0 > \; ,
\end{equation}
where $| 0 >$ is the vacuum, and $N' = \frac{N}{2}$ is the number of pairs. Now, from the identity~\cite{Anderson:1966}: 
\begin{equation}
\label{3.4} 
 | \Omega >_N  \; = \;  \frac{1}{2\pi} \;   \int_0^{2 \pi}  d \phi  \; e^{- i N' \phi} \;  \exp{[ e^{i \phi } { \it{b}^{\dagger}}]}  \; | 0 >  \; ,
\end{equation}
we argue that the state:
\begin{equation}
\label{3.5} 
 | \Omega >_{BCS}  \; = \;   \exp{[ e^{i \phi } { \it{b}^{\dagger}}]}  \; | 0 >  
\end{equation}
is the superconducting ground state for a system with varying number of particles. Obviously, in that case the relevant Hamiltonian is $\hat{H} - \mu \hat{N}$, where
$\mu$ is the chemical potential. From Eq.~(\ref{3.2}) we obtain:
\begin{equation}
\label{3.6} 
{ \it{b}^{\dagger}} \; = \; \sum_{\vec{q}}   \;  \varphi(\vec{q}) \;    \hat{\it{c}}^{\dagger}_{\vec{q},\uparrow}   \;   \hat{\it{c}}^{\dagger}_{-\vec{q},\downarrow}  \; ,
\end{equation}
where:
\begin{equation}
\label{3.7} 
 \varphi(\vec{r}_1-\vec{r}_2) \;  = \;    \sum_{\vec{q}}  \exp{[ i \vec{q} \cdot (\vec{r}_1-\vec{r}_2)]}   \;  \varphi(\vec{q}) \; .
\end{equation}
Using   Eq.~(\ref{3.6}) it is easy to find:
\begin{equation}
\label{3.8} 
 | \Omega >_{BCS}  \; =  \;   \prod_{\vec{q}}  \;  {\cal{N}}_{\vec{q}}  \; \left [ 1  \;  + \;   e^{i \phi } \;  \varphi(\vec{q}) \;         \hat{\it{c}}^{\dagger}_{\vec{q},\uparrow}   \;   
 \hat{\it{c}}^{\dagger}_{-\vec{q},\downarrow}  \right ]\;  |0>  \; ,  
\end{equation}
with the normalization factor $ {\cal{N}}_{\vec{q}}$ . Indeed, Eq.~(\ref{3.8}) agrees with the well-known BCS variational ground state:
\begin{equation}
\label{3.9} 
 | O >_{BCS}  \; =  \;   \prod_{\vec{q}}  \;  \left [   u(\vec{q})    \;  + \;    v(\vec{q})     \;   \hat{\it{c}}^{\dagger}_{\vec{q},\uparrow}   \;   
 \hat{\it{c}}^{-\dagger}_{-\vec{q},\downarrow}  \right ]\;  |0>  \; ,  
\end{equation}
with the normalization:
\begin{equation}
\label{3.10} 
  | u(\vec{q})|^2    \;  + \;   | v(\vec{q}) |^2   \; = \; 1 \; .  
\end{equation}
As we discuss later on, the superconducting ground-state wave function in the form of Eq.~(\ref{3.1}) will be of relevance in the underdoped region, while in the overdoped region the BCS variational ground state Eq.~(\ref{3.9}) would be more appropriate.
\subsection{\normalsize{Real space d-wave bound states}}
\label{3-1}
We are interested in the problem of two holes in interaction with the effective two-body potential  Eq.~(\ref{2.7}). The resulting Schr\"odinger equation is:
\begin{equation}
\label{3.11} 
 \left [  - \, \frac{ \hbar^2}{2 m^*_h}  \left (   \nabla^2_{ \vec{r}_1} \; + \; \nabla^2_{\vec{r}_2}   \right ) + V(\vec{r}_1 - \vec{r}_2 ) \right ] \; \Phi( \vec{r}_1  - \vec{r}_2 ) = \; E \; 
 \Phi (\vec{r}_1 - \vec{r}_2 )    \; .  
\end{equation}
According to  the reduced interaction Hamiltonian  Eq.~(\ref{2.6}) the holes have antiparallel spins, so that the wavefunction $\Phi( \vec{r}_1  - \vec{r}_2 )$ must be symmetric under hole interchange. Introducing:
\begin{equation}
\label{3.12} 
\vec{r} \; = \; \vec{r}_1  - \vec{r}_2  \; \; \; \; , \; \; \; \; \vec{R} \; = \frac{\vec{r}_1  + \vec{r}_2}{2} \; ,
\end{equation}
we rewrite  Eq.~(\ref{3.11}) as:
\begin{equation}
\label{3.13} 
 \left [  - \, \frac{ \hbar^2}{2 m^*_h}  \left (  2 \, \nabla^2_{ \vec{r}} \; + \; \frac{1}{2} \, \nabla^2_{\vec{R}}   \right ) + V(r) \right ] \; \Phi( \vec{R} ,    \vec{r} ) = \; E \; 
 \Phi (\vec{R} ,  \vec{r} )    \; ,
\end{equation}
whereupon:
\begin{equation}
\label{3.14} 
 \Phi (\vec{R} ,  \vec{r} )   \; =  \;    \exp{ (i  \vec{K}  \cdot  \vec{R})} \; \varphi(  \vec{r} )            \; ,
\end{equation}
\begin{equation}
\label{3.15} 
E  \; =  \;   \frac{\hbar^2 \, \vec{K}^2}{4 \, m^*_h}    \; - \; \Delta             \; ,
\end{equation}
\begin{equation}
\label{3.16} 
 \left [  - \, \frac{ \hbar^2}{ m^*_h}    \nabla^2_{ \vec{r}} \;  + \; V(r) \right ] \; \varphi( \vec{r}  ) = - \; \Delta  \;   \varphi( \vec{r}  )    \; .
\end{equation}
Without loss in generality, we may assume that the center of mass of the pair is at rest, $\vec{P}= \hbar \, \vec{K} = 0$. Since the potential is central we may adopt
polar coordinates. We have:
\begin{equation}
\label{3.17} 
\varphi( r , \theta  )  \; =  \;   \frac{\exp{( i m \theta)} }{\sqrt{2 \pi}}  \;    \varphi_m( r )         \; ,
\end{equation}
Inserting into  Eq.~(\ref{3.16}) and using  Eq.~(\ref{2.7}) we get:
\begin{equation}
\label{3.18} 
\varphi_m(r=0)  \; =  \;  0 \; , 
\end{equation}
\begin{equation}
\label{3.19} 
 \left [  \frac{d^2}{d r^2 }  + \frac{1}{r} \,   \frac{d}{d r } \; - \;  \frac{m^2}{r^2} \; + \; 
\frac{ m^*_h}{ \hbar^2} \left ( - \Delta_m \, + \, V_0 \right )   \right ] \;   \varphi_m( r )  =  \;  0    \; \;  , \; \;  0 \; < \; r \; < \; r_0(\delta) \; ,
\end{equation}
\begin{equation}
\label{3.20} 
 \left [  \frac{d^2}{d r^2 }  + \frac{1}{r} \,   \frac{d}{d r } \; - \;  \frac{m^2}{r^2} \; - \; 
\frac{ m^*_h}{ \hbar^2}  \Delta_m  \right ] \;   \varphi_m( r )  =  \;  0     \; \;  , \; \; r_0(\delta) \; < \; r \;  \; .
\end{equation}
Since  $\varphi_m( r )$ is even and must satisfy  Eq.~(\ref{3.18}), we have $ m = 2, 4, ...$. With our choice of the parameters, Eqs.~(\ref{2.8}), (\ref{2.9}) and
 (\ref{2.10}), it is straightforward to check  that there are bound states ($\Delta_m > 0$) only for $m=2$ (d-wave). In this case we get:
\begin{equation}
\label{3.21} 
  \varphi_2( r )   \;  = \; 
 \left\{ \begin{array}{ll}
 \; \; A \; J_2(\tilde{k} r)  \; \; \; \; &   r \; < \; r_0(\delta) 
  \\
 \; \; B \; K_2(\tilde{\kappa} r)  \; \; \; \; &    r_0(\delta)  \;  <  \; r
 \end{array}
    \right.
\end{equation}
where $J_2(x)$ and $K_2(x)$ are Bessel functions,  
\begin{equation}
\label{3.22} 
\tilde{\kappa}^2  \; =  \;  \frac{ m^*_h}{ \hbar^2}  \Delta_2 \;  \; \; , \; \; \;  \tilde{k}^2  \; =  \;  \frac{ m^*_h}{ \hbar^2}  \; (V_0 \; - \;  \Delta_2) \; ,
\end{equation}
and
\begin{equation}
\label{3.23} 
 \tilde{\kappa} r_0(\delta)  \;  \; \frac{ K^{'}_2(\tilde{\kappa} r_0(\delta) )}{ K_2(\tilde{\kappa} r_0(\delta) )}    \; =  \; 
 \tilde{k} r_0(\delta)  \;  \; \frac{ J^{'}_2(\tilde{k} r_0(\delta) )}{ J_2(\tilde{k} r_0(\delta) )}   \; . 
\end{equation}
With the further definitions:
\begin{equation}
\label{3.24} 
x_0 \; = \; \tilde{k}_0 \; r_0(\delta) \; \; , \; \; \tilde{k}_0^2 \; = \; \frac{ m^*_h V_0}{ \hbar^2}  \;  \;  , \; \;  \zeta \; = \; \frac{\tilde{k}}{\tilde{k}_0} \; , 
\end{equation}
one can rewrite  Eq.~(\ref{3.23}) as:
\begin{equation}
\label{3.25} 
  x_0 \; \sqrt{1-\zeta^2}  \;  \; \frac{ K^{'}_2(  x_0 \; \sqrt{1-\zeta^2}   )}{ K_2( x_0 \; \sqrt{1-\zeta^2} )}    \; =  \; 
  x_0 \; \zeta  \;  \; \frac{ J^{'}_2( x_0 \; \zeta  )}{ J_2( x_0 \; \zeta  )}   \; . 
\end{equation}
It is not too hard to see that Eq.~(\ref{3.25}) allows non trivial solutions if the {\it effective} wall-depth $x_0$ satisfies: 
\begin{equation}
\label{3.26}  
  x_0 \; \geq \;   \overline x_0 \; \; \; , \; \; \;  \overline x_0 \; \simeq \; 3.8317 \; ,
\end{equation}
where $  \overline x_0 $ is the solution of:
\begin{equation}
\label{3.27} 
  x \; \frac{ J^{'}_2( x  )}{ J_2( x  )} \; + \; 2 \; = \; 0 \; .   
\end{equation}
\begin{figure}[t]
\begin{center}
\includegraphics[width=0.8\textwidth,clip]{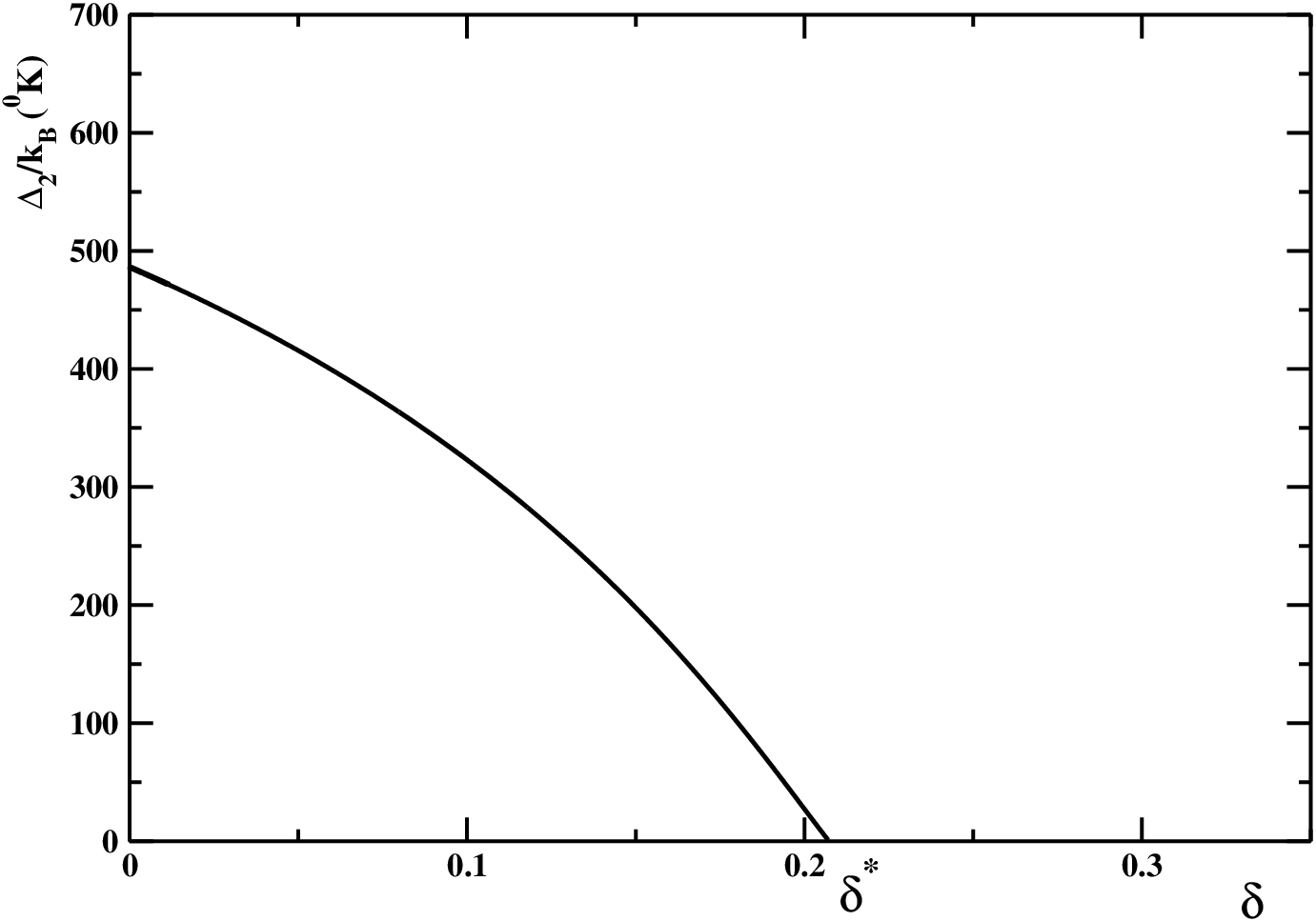}
\caption{\label{fig-3}  The gap $\Delta_2(\delta)$ versus the hole doping fraction $\delta$.  According to Eq.~(\ref{3.28}) $\delta^* \simeq 0.207$.  }
\end{center}
\end{figure}
From Eq~(\ref{3.26}) it follows that there are non trivial solutions for hole doping such that:
\begin{equation}
\label{3.28}  
  0 \; \leq \;  \delta \; \leq \delta^* \; \; \; , \; \; \; \delta^* \; \simeq \; 0.207 \; . 
\end{equation}
In fact, in Fig~\ref{fig-3} we display $\Delta_2(\delta)$ versus the hole doping fraction $\delta$. We see that the gap $\Delta_2(\delta)$ decreases rapidly with
$\delta$ and vanishes at $\delta=\delta^*$ according to Eq.~(\ref{3.28}). Note that, with our parameters, we have $\Delta(0) \simeq 486 \; K  \cdot  k_B$ which
should be  compared with $V_0 \simeq 1021 \;  K  \cdot k_B$.
\begin{figure}[t]
\begin{center}
\includegraphics[width=0.8\textwidth,clip]{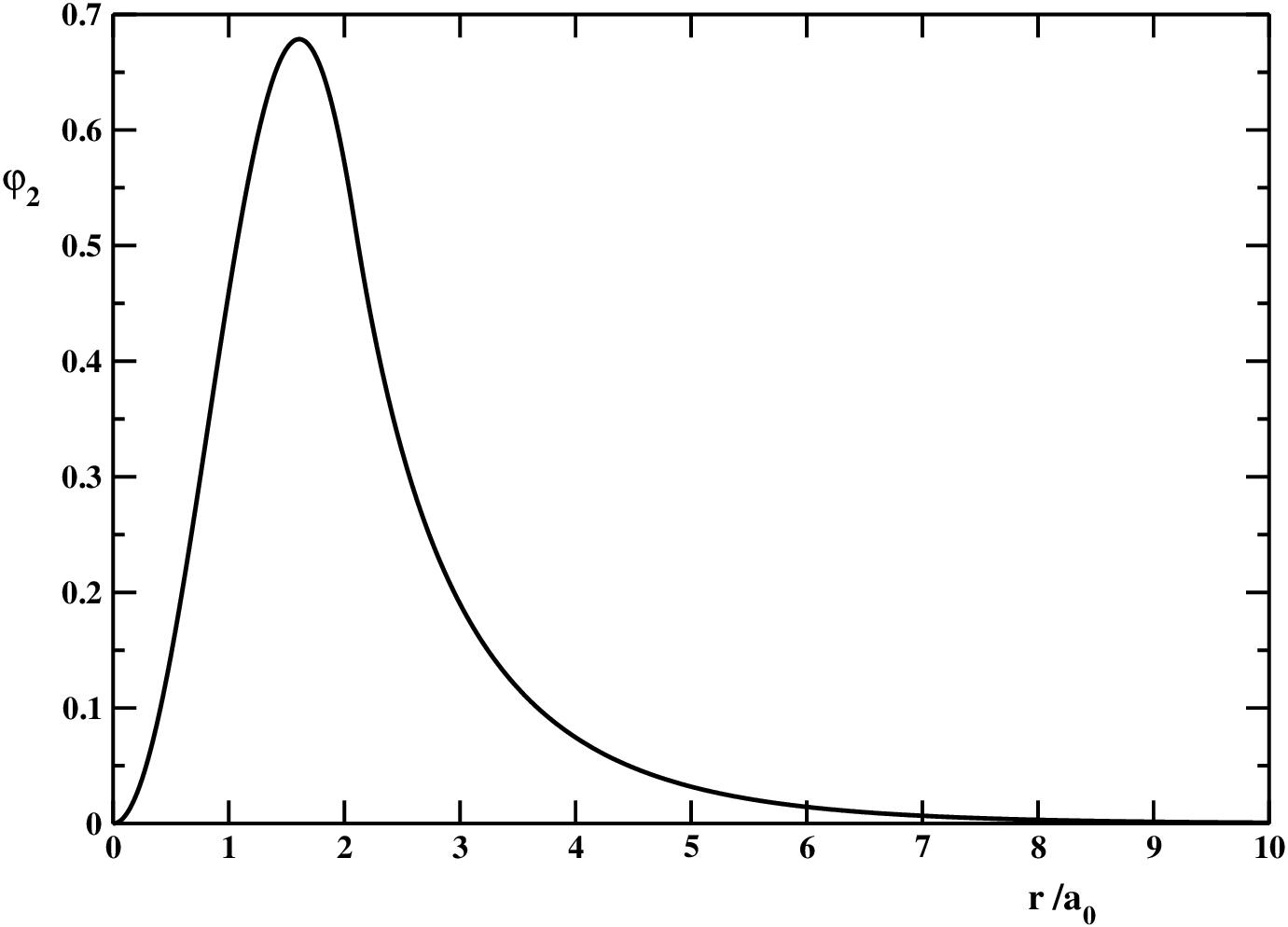}
\caption{\label{fig-4}  The normalized d-wave wavefunction $\varphi_2$ for $\delta = 0.18$. }
\end{center}
\end{figure}
From Figure~\ref{fig-4}, where we  display the normalized wavefunction   $\varphi_2( r )$,  Eq.~(\ref{3.21}), for a typical value of $\delta$, we see that our bound-state
wavefunction extends up to distance $ r \sim 6 a_0$. \\
Since the eigenvalue equations  Eqs.~(\ref{3.19}) and (\ref{3.20}) depend on $m^2$, the general d-wave wavefunction is a linear combination of $\varphi_2( r )$ 
and $\varphi_{-2}( r )$. From the geometry of the $CuO_2$ planes we are lead to assume:
\begin{equation}
\label{3.29}  
\varphi_2( r,\theta) \; = \;   \varphi_2( r) \cdot \cos{(2 \theta)}    \; , 
\end{equation}
where the coordinate axis are directed along the $Cu-O$ bond directions as in Fig.~\ref{fig-2}. This is the most natural choice since the wavefunction is sizable along the $Cu-O$ bonds and vanishes at $\theta = \pm \frac{\pi}{4}$. For later convenience, we need  to evaluate the Fourier transform of the wavefunction  Eq.~(\ref{3.29}). We have: 
\begin{equation}
\label{3.30}  
\tilde{\varphi}_2( \vec{k}) \; = \;  \int d \vec{r} \;  \varphi_2( r,\theta) \; \exp{ (i \vec{k} \cdot \vec{r}) }   \; . 
\end{equation}
Assuming that the $k_x$, $k_y$ axis are oriented along the copper-oxygen bond directions, and using the expansion:
\begin{equation}
\label{3.31}  
\exp{ (i \vec{k} \cdot \vec{r}) } \; = \;  \exp{ (i k r \cos{\theta_{kr} } )}  \; = \; \sum_{m=-\infty}^{+ \infty}  \; \exp{ [i m (\theta_{kr} + \frac{\pi}{2} )] }   \; J_m( kr)   \; ,
\end{equation}
we readily obtain:
\begin{equation}
\label{3.32}  
\tilde{\varphi}_2(k, \theta_k ) \; = \; \tilde{\varphi}_2(k) \,  \cos{(2 \theta_k)} 
\end{equation}
where:
\begin{equation}
\label{3.33}  
\tilde{\varphi}_2(k) \; = \; -  2 \pi  \int_0^{\infty} d r  \, r \,   \varphi_2(r) \, J_2(kr) \; . 
\end{equation}
Remarkably, we see that the Fourier transform of the wavefunction vanishes along the {\it nodal} directions:
\begin{equation}
\label{3.34}  
 \theta_k \; = \;  \pm \; \frac{\pi}{4} \; , 
\end{equation}
while it is sizable along the {\it antinodal} directions:
\begin{equation}
\label{3.35}  
 \theta_k \; = \;  0 \; , \; \pm \; \frac{\pi}{2}  \; . 
\end{equation}
These features are reminiscent of the observed pseudogap in the underdoped region. Therefore, we are lead to identify $\Delta_2(\delta)$ with the pseudogap
and to introduce the pseudogap temperature according to:
\begin{equation}
\label{3.36}  
k_B \, T^*(\delta) \; = \;  \frac{1}{2} \, \Delta_2(\delta) \;  . 
\end{equation}
\subsection{\normalsize{Superconductivity in the underdoped region and the pseudogap}}
\label{3-2}
We have shown in the previous Section that, as long as $\delta < \delta^*$, it is energetically favored to pair two holes into a d-wave bound state with wavefunction given by  Eq.~(\ref{3.29}). Then, we may construct the superconducting ground state according to Eq.~(\ref{3.1}) with $\varphi( \vec{r}_i - \vec{r}_j) = \varphi_2( \vec{r}_i - \vec{r}_j)$.
However, although the formation of Cooper pairs is essential in forming the superconducting state, its remarkable properties, i.e. zero resistance and perfect diamagnetism, require phase coherence among the pairs. In the underdoped region $\delta \ll \delta^*$ we infer from Fig.~\ref{fig-4} that the size of the pair is small with respect to the average distance between holes. So that  the ground state wavefunction Eq.~(\ref{3.1}) reduces to the ground state of a Bose-Einstein gas of paired holes. In this case the required phase coherence is established by the condensation of pairs. However, in two spatial dimensions it is well known that there is no Bose-Einstein condensation. Thus we are lead to conclude that in the extremely underdoped region $\delta \ll \delta^*$ there is a pseudogap $\Delta_2(\delta)$ but not superconductivity. However, by increasing $\delta$ it will be a certain minimal value of the hole doping fraction $\delta_{min}$ where the pairs become to overlap. To estimate $\delta_{min}$, we observe that the pair wavefunction
$\varphi_2(r)$ extends up to distance $r \sim \xi_0 \simeq 6 a_0$. Thus,  to have overlap between different hole pairs we need at least two holes in a square of size  
$6 a_0$. Thereby we estimate:
\begin{equation}
\label{3.37}  
\delta_{min}  \; \simeq  \;  \frac{2}{6^2} \; \simeq \; 0.056   \;  , 
\end{equation}
which agrees, remarkably, with the observed value $\delta_{min} \simeq 0.05 - 0.06$ for the onset of the superconductivity in the hole-doped cuprates.  Once the pairs
begin to overlap, the phases of the pairs are locked to a constant value. Indeed, it is well known~\footnote{ For instance, see Ref.~\cite{Legget:2008}.} that a condensate
with varying phase $\Theta(\vec{R})$ contribute to the ground state energy with:
\begin{equation}
\label{3.38}  
H_{ \Theta}  \; =  \; \frac{K_s}{2} \; \int d\vec{R} \, | \nabla \Theta(\vec{R}) |^2 \;  , 
\end{equation}
where $K_s$ is the so-called phase stiffness. In our case we have:
\begin{equation}
\label{3.39}  
K_s \; \simeq \;  \frac{\hbar^2 }{2 m^*_h}  \; n_s
\end{equation}
with the superfluid density given by:
\begin{equation}
\label{3.40}  
 n_s \; \simeq \frac{\delta}{2 a^2_0} \; \; , \; \; \delta_{min} \leq \delta \; .
\end{equation}
\begin{figure}[t]
\begin{center}
\includegraphics[width=0.8\textwidth,clip]{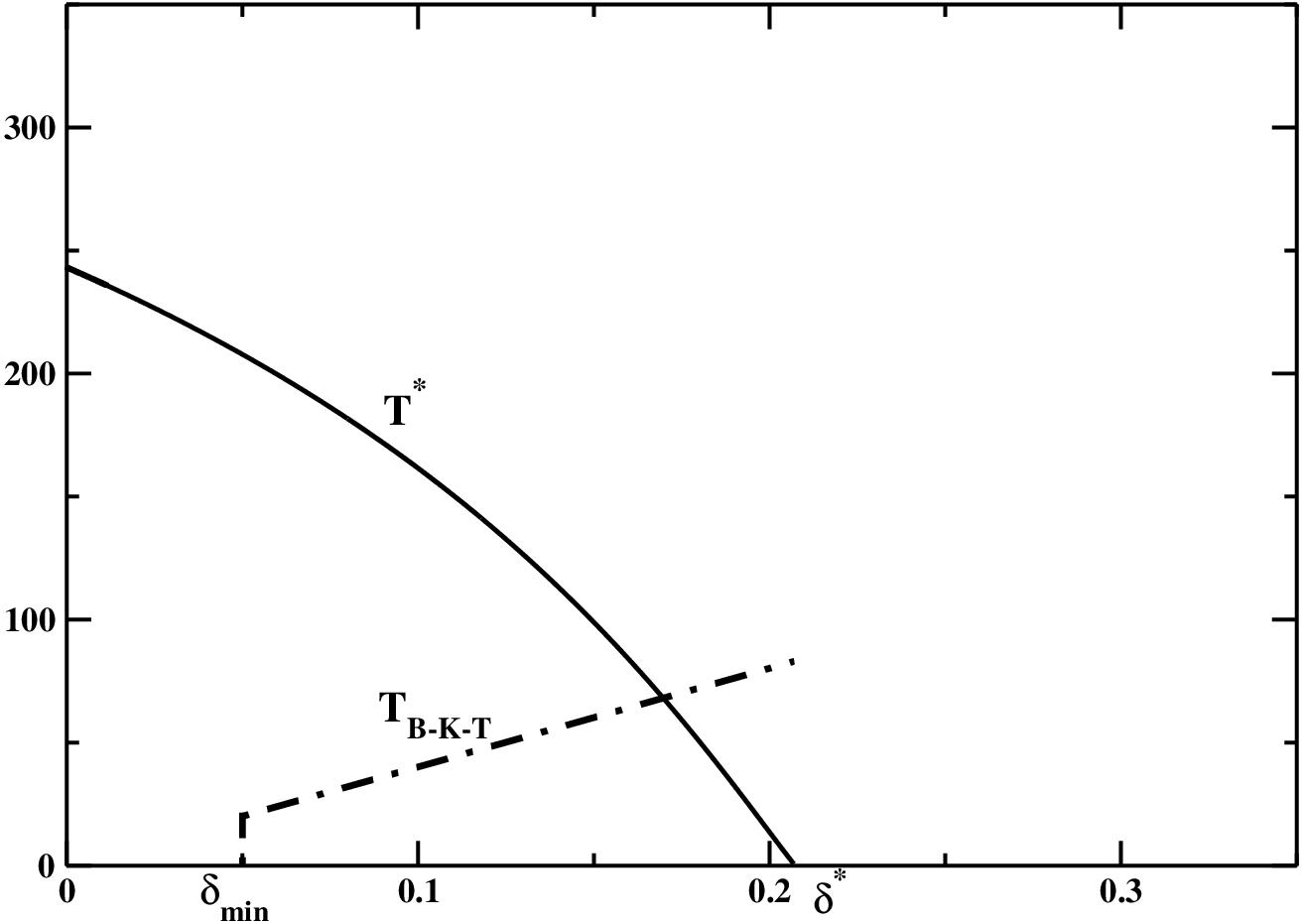}
\caption{\label{fig-5}  The pseudogap  and  Berezinskii-Kosterlitz-Thouless temperatures (in Kelvin) versus  $\delta $.
We assume $\delta_{min} \simeq 0.05$.}
\end{center}
\end{figure}
As is well known~\cite{Berezinskii:1971,Kosterlitz:1973}, the phase coherence of the condensate survives up to the Berezinskii-Kosterlitz-Thouless (B-K-T) critical
temperature:
\begin{equation}
\label{3.41}  
 k_B  \; T_{B-K-T} \; \simeq \; \frac{\pi}{2} \;  K_s(T_{B-K-T})  \; .
\end{equation}
For temperatures above $T_{B-K-T}$ the phase coherence and the superconductivity is lost due to the thermal activation of vortex excitations. If we neglect the
temperature dependence of the phase stiffness, we obtain:
\begin{equation}
\label{3.42}  
 k_B  \; T_{B-K-T} \; \simeq \; \frac{\pi}{8} \;  \frac{\hbar^2 }{ m^*_h} \;  \frac{\delta}{ a^2_0} \;  \simeq \;  \pi \,   \frac{t^2 }{ U} \; \delta \; \; , \; \; \delta_{min} \leq \delta \; .
\end{equation}
Using our numerical values for $t$ and $U$ we get:
\begin{equation}
\label{3.43}  
T_{B-K-T} \; \simeq \; 401 \; K \; \; \delta \; \; , \; \; \delta_{min} \leq \delta \; .
\end{equation}
In Figure~\ref{fig-5} we display the Berezinskii-Kosterlitz-Thouless critical temperature together with the pseudogap temperature $T^*$  versus
the hole doping fraction by assuming $\delta_{min} \simeq 0.05$. We see that in our approach the superconductivity transition  in the underdoped region is
described by the $B-K-T$ transition at $T_c=T_{B-K-T}$, which coexists with the much higher pseudogap temperature $T^*$. This picture is supported by a large amounts  of observational data which, however, will be discussed in details in Sect.~\ref{4}. 
\subsection{\normalsize{d-wave BCS ground state}}
\label{3-3}
To complete the phase diagram of the hole-doped cuprate superconductors we must address the problem of the ground state for $\delta > \delta^*$.
Since the pseudogap $\Delta_2(\delta)$ vanishes at $\delta = \delta^*$, we do not have real-space paired holes and, whence, the $B-K-T$ critical
temperature cannot be defined. On the other hand, as long as $\delta < \delta_c \simeq 0.35$, the attractive two-body potential Eq.~(\ref{2.7}) is short-range and can be
considered as a small perturbation. As a consequence, in the overdoped region $\delta > \delta^*$ the normal state is described by the Hamiltonian $\hat{H}_0$, Eq.~(\ref{2.18}).
In other words, the normal state properties are those of ordinary Fermi liquid metals that is characterized by hole quasiparticles with effective mass given by Eq.~(\ref{2.17}) and density:
\begin{equation}
\label{3.44}  
n_h \; \simeq \;  \frac{1 + \delta}{a_0^2}  \; \;  .
\end{equation}
In fact, in the overdoped side of the superconducting phase diagram, experiments strongly suggest that  above
the critical temperature the full Fermi surface is restored with well-defined quasiparticles. Thereby this confirms the applicability of a Fermi liquid picture.
Moreover, quantum oscillation experiments~\cite{Vignolle:2008} in an overdoped high-temperature superconductor are consistent with
hole quasiparticles with density given by Eq.~(\ref{3.44}) and effective mass:
\begin{equation}
\label{3.45}  
m^*_h  \; = \;  ( 4.1 \pm 1.0 ) \; m_e 
\end{equation}
in fair agreement with our estimate, Eq.~(\ref{2.19}). \\
The superconductive instability is driven by the short-range attractive interaction between the quasiparticles. We see, then, that the pairing is
in momentum space, so that the relevant superconducting ground state is the BCS variational ground state Eq.~(\ref{3.9}) with Hamiltonian
$\hat{H}-\mu \hat{N}$, where the chemical potential is essentially the Fermi energy:
\begin{equation}
\label{3.46}  
\mu \; \simeq \; \varepsilon_F \; \; , \; \;  \varepsilon_F \; = \; \frac{\hbar^2 k_F^2}{2 m^*_h} \; \; , \; \; k_F \; \simeq \; \sqrt{2 \pi n_h} \; .
\end{equation}
The relevant gap equation has been discussed since long time~\cite{Anderson:1961}:
\begin{equation}
\label{3.47}  
\Delta(\vec{k}) \; = \; - \frac{1}{2} \; \sum_{\vec{k}'} \; \; \frac{ V(\vec{k} - \vec{k}') \; \Delta(\vec{k}')} { \sqrt{ \xi^2_{\vec{k}'} + |  \Delta(\vec{k}')|^2 } } 
\end{equation}
where:
\begin{equation}
\label{3.48}  
\xi_{\vec{k}} \;  =  \; \frac{\hbar^2 \vec{k}^2}{2 m^*_h} \; -  \; \varepsilon_F 
\end{equation}
and 
\begin{equation}
\label{3.49}  
V(\vec{k} - \vec{k}')  \;  =  \;  \int d \vec{r} \; \exp{ [- i ( \vec{k} - \vec{k}') \cdot \vec{r}]} \; V(r) \; .
\end{equation}
Using the expansion Eq.~(\ref{3.31}) we may recast  Eq.~(\ref{3.47}) into:
\begin{equation}
\label{3.50}  
\Delta(\vec{k}) \;  = \; - \frac{1}{2} \sum_{\vec{k}'} \; \sum_{m=-\infty}^{+\infty}  \exp{(i m \theta_{kk'})} \; 
 \; \frac{ V_m(k, k')  \Delta(\vec{k}')} { \sqrt{ \xi^2_{\vec{k}'} + |  \Delta(\vec{k}')|^2 } }  \; ,
\end{equation}
with:
\begin{equation}
\label{3.51}  
V_m(k, k')  \;  =  \;  2 \pi \;  \int_0^{\infty}  dr \; r \; V(r) \; J_m(kr) \; J_m(k'r)  \; .
\end{equation}
Equation~(\ref{3.50}) suggests that:
\begin{equation}
\label{3.52}  
\Delta(\vec{k}) \;  = \; \Delta(k,\theta_k) \;  = \; \sum_{\ell = - \infty}^{+\infty}  \exp{(i \ell \theta_{k})} \;  \Delta_{\ell}(k)   \; ,
\end{equation}
In the weak coupling limit one expects that the mixing of different partial wave $\ell$ will be small~\cite{Anderson:1961}. Therefore Eq.~(\ref{3.50}) reduces to:
\begin{equation}
\label{3.53}  
\Delta_m(k,\theta_k) \; \simeq \; - \frac{1}{2} \sum_{\vec{k}'} \;   \exp{(i m \theta_{kk'})} \; 
 \; \frac{ V_m(k, k')  \Delta_m(k',\theta_{k'})} { \sqrt{ \xi^2_{\vec{k}'} + |  \Delta_m(k',\theta_{k'})|^2 } }  \; .
\end{equation}
Since the BCS gap in the limit of localized pairs is proportional to the bound-state wave function in momentum space, we conclude that Eq.~(\ref{3.53})
admits non trivial solutions for the d-wave gap ($m= \pm 2$). Moreover, from Eq.~(\ref{3.32}) we infer that:
\begin{equation}
\label{3.54}  
\Delta_2(k,\theta_k) \; \equiv \; \Delta_{BCS}(k,\theta_k) \; = \;  \Delta_{BCS}(k) \; \cos{(2 \theta_k)} \; .
\end{equation}
Thus, we obtain the following gap equation:
\begin{equation}
\label{3.55}  
 \Delta_{BCS}(k) \;  \simeq \; -  \sum_{\vec{k}'} \;  [\cos{(2 \theta_{k'})]^2} \; 
 \; \frac{ V_2(k, k')  \Delta_{BCS}(k')} { \sqrt{ \xi^2_{\vec{k}'} + [  \Delta_{BCS}(k') \cos{(2 \theta_{k'})}]^2 } }  \; .
\end{equation}
Since the gap is sizable on the Fermi surface, we may further simplify Eq.~(\ref{3.55}) as:
\begin{equation}
\label{3.56}  
1 \;  \simeq \; -   V_2  \; \int \frac{d\vec{k}'}{(2 \pi)^2}  \; 
 \; \frac{  [\cos{(2 \theta_{k'})}]^2 } { \sqrt{ \xi^2_{\vec{k}'} + [  \Delta_{BCS} \cos{(2 \theta_{k'})}]^2 } }  
\end{equation}
where $\Delta_{BCS} =  \Delta_{BCS}(k_F)$ and $V_2= V_2(k_F,k_F)$. A standard calculation gives:
\begin{equation}
\label{3.57}  
1 \;  \simeq \; -   \frac{V_2 m^*_h}{\hbar^2}  \; \int_{-\varepsilon_c}^{+\varepsilon_c}  \frac{d \xi}{(2 \pi)^2}  \;  \int_{0}^{2 \pi} d\theta
 \; \frac{  [\cos{(2 \theta)}]^2 } { \sqrt{ \xi^2 + [  \Delta_{BCS} \cos{(2 \theta)}]^2 } }  
\end{equation}
where $\varepsilon_c$ is an energy cut-off much smaller than the Fermi energy. Performing the integrals and using the approximation:
\begin{equation}
\label{3.58}  
archsinh \left [   \frac{\varepsilon_c}{ \Delta_{BCS} |\cos{(2 \theta)}| }   \right ]  \;   \simeq  \;   
 \ln{  \left  [  \frac{2 \varepsilon_c}{  \Delta_{BCS} |\cos{(2 \theta)| } }\right ] } \; ,
\end{equation}
we get:
\begin{equation}
\label{3.59}  
\Delta_{BCS} \;  \simeq \;   \frac{4 \varepsilon_c}{\sqrt{e}}   \;  \exp{ [ -  \frac{1}{\lambda_2}]}  \; ,
\end{equation}
where:
\begin{equation}
\label{3.60}  
\lambda_2  \; = \; \frac{m^*_h V_0}{\hbar^2} \; \int_{0}^{r_0(\delta)} dr \; r \left [ J_2(k_F r) \right ]^2 \; .
\end{equation}
\begin{figure}[t]
\begin{center}
\includegraphics[width=0.8\textwidth,clip]{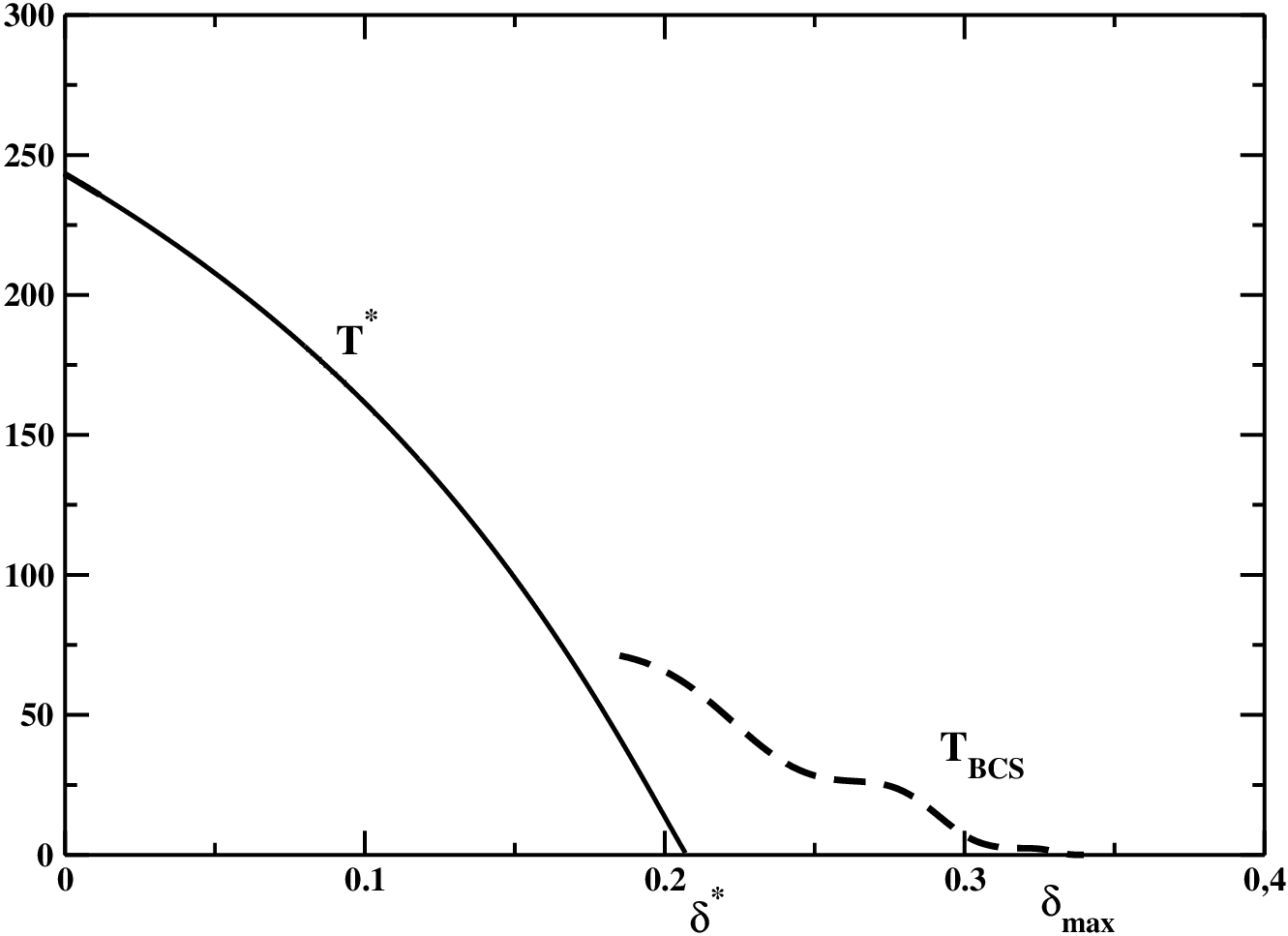}
\caption{\label{fig-6}  The pseudogap and  BCS critical  temperatures (in Kelvin) versus the hole doping fraction. }
\end{center}
\end{figure}
To obtain the critical temperature, we note that~\cite{Won:1994}:
\begin{equation}
\label{3.61}  
\frac{\Delta_{BCS}} {k_B T_c} \; \simeq \; \pi e^{\ln{2} - \frac{1}{2} - \gamma}  \; ,
\end{equation}
$ \gamma  = 0.577216...$ being the Euler's constant.  As concern the high-energy cut-off, it is natural to identify $\varepsilon_c$ with $\Delta_2(\delta=0)$ which,
indeed, is much smaller than the Fermi energy in the range of hole doping fraction of interest. In summary, we obtain for the critical temperature $T_c \equiv T_{BCS}$:
\begin{equation}
\label{3.62}  
k_B \; T_{BCS} \;  \simeq \;   \frac{2  e^{\gamma}}{\pi}   \;  \Delta_2(0) \;  \exp{ [ -  \frac{1}{\lambda_2}]}  \; .
\end{equation}
Equation~(\ref{3.62}) shows that the BCS critical temperature depends on the hole doping fraction. In fact, in Fig.~\ref{fig-6} we compare $T_{BCS}(\delta)$ with
the pseudogap temperature $T^*(\delta)$. It is evident that for $\delta^* \leq \delta \leq \delta_{max} \simeq 0.30$ we have a non-zero BCS d-wave gap. Note that
the undulations on the critical temperature curve is an artifact of the sharp cut-off of the effective interaction potential at $r=r_0(\delta)$.
\subsection{\normalsize{Phase diagram of hole-doped cuprates}}
\label{3-4}
\begin{figure}[t]
\begin{center}
\includegraphics[width=0.8\textwidth,clip]{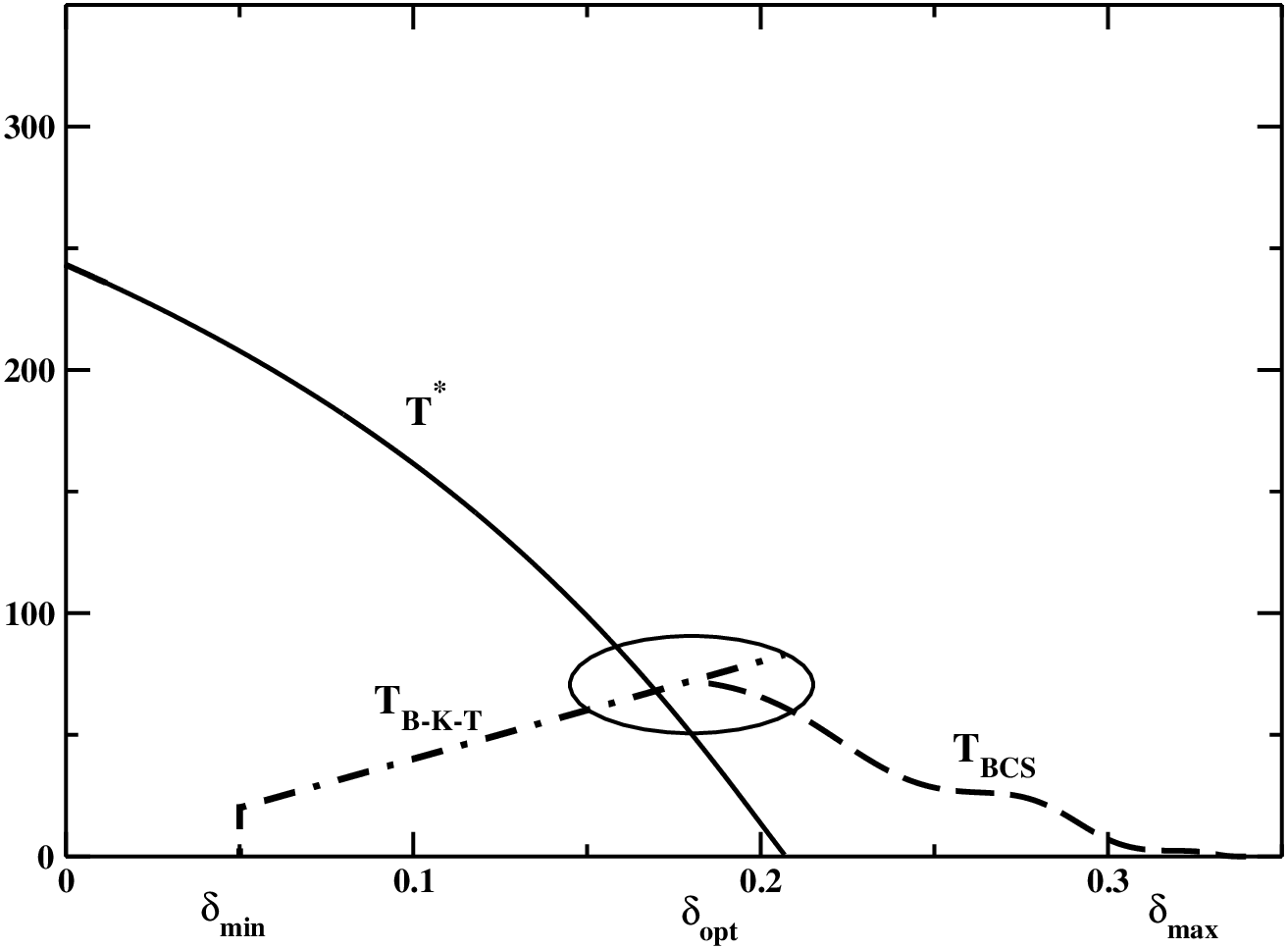}
\caption{\label{fig-7}  The pseudogap, Berezinskii-Kosterlitz-Thouless, and  BCS critical temperatures (in Kelvin) versus  the hole doping fraction $\delta $. 
The region enclosed by the ellipse where the three temperatures are comparable is the optimal doped region. }
\end{center}
\end{figure}

We are, now, in position of discuss the phase diagram of the hole-doped cuprates. In Fig.~\ref{fig-7} we display the Berezinskii-Kosterlitz-Thouless ($T_{B-K-T}$), the pseudogap ($T^*$), and the BCS ($T_{BCS}$) temperatures versus the hole doping fraction $\delta$.
In our approach the superconducting dome-shaped region of the phase diagram of oxide cuprates is determined by the region enclosed by the two critical temperatures 
$T_{B-K-T}$ and $T_{BCS}$. The Berezinskii-Kosterlitz-Thouless temperature increases with the doping $\delta$, while the BCS critical temperature decreases with increasing hole doping. These temperatures are comparable for $\delta \sim \delta_{opt}$. Accordingly, we individuate three regions in the phase diagram:
\begin{itemize}
\item the underdoped region   $\delta_{min} \; \lesssim \; \delta \; < \; \delta_{opt} $  ;
\item  the optimal doped region   $ \delta \; \thickapprox \; \delta_{opt}$ ;
\item the overdoped region      $ \delta_{opt} \; < \; \delta \;  \lesssim  \;  \delta_{max}$  .
\end{itemize}
In  Fig.~\ref{fig-7} we have also displayed the pseudogap temperature $T^*(\delta)$ which decreases rapidly with increasing doping, meets the superconducting dome
in the optimal doped region, and vanishes beyond a critical doping level $\delta^*$ within the optimal doped region. In the following Sections we shall discuss the 
physical properties which characterize the three regions and compare with selected experimental observations supporting our proposal.
\section{The physics of the underdoped region}
\label{4}
\begin{figure}[t]
\begin{center}
\includegraphics[width=0.8\textwidth,clip]{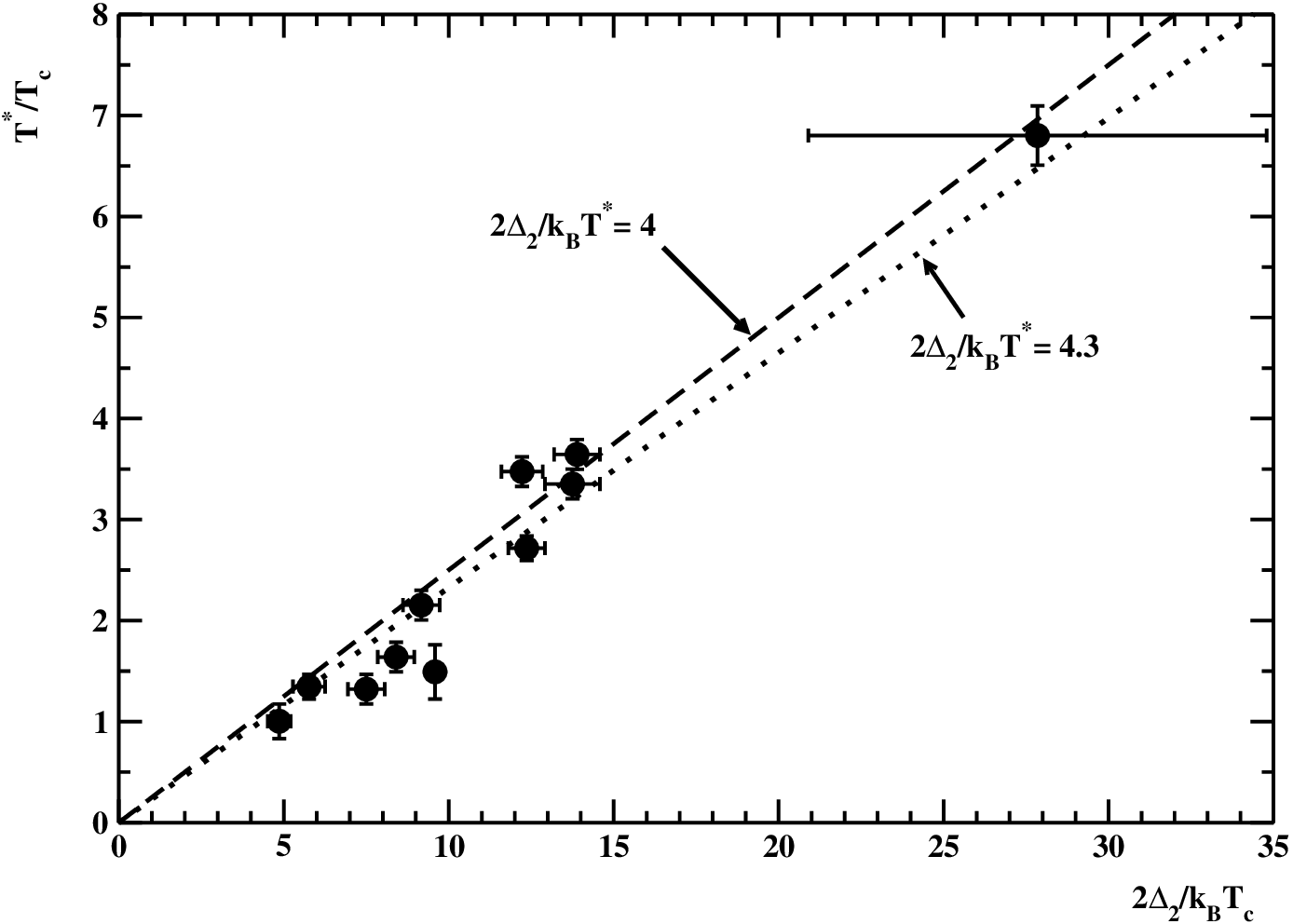}
\caption{\label{fig-8}  $\frac{T^*}{T_c}$ versus $\frac{2 \Delta_2}{k_B T_c}$ for various cuprates compared to our relation  $\frac{2 \Delta_2}{k_B T^* } = 4$ (dashed line) and
the d-wave BCS relation  $\frac{2 \Delta_2}{k_B T^* } = 4.3$ (dotted line). The data have been extracted from Fig.~5 of Ref.~\cite{Kugler:2001}.  }
\end{center}
\end{figure}

The underdoped region in all hole-doped high-temperature superconductors is characterized by the presence of the pseudogap~\footnote{ For a review on the pseudogap in cuprates,  see Refs.~\cite{Timusk:1999,Tallon:2001}.}. 
In general it is assumed that the pseudogap has a d-wave structure. Moreover, there is a general
 agreement on the  doping dependence of the pseudogap. For instance, recently the authors of
Ref.~\cite{Chatterjee:2010} used angle resolved photoemission spectroscopy to probe the electronic excitations in underdoped $Bi_2Sr_2CaCu_2O_{8+\delta}$. These authors find evidence of a nodal liquid whose excitation gap vanishes only at points in momentum space which are consistent with the gap structure of a d-wave superconductors. In particular, these
authors show that the measured gap in momentum space is consistent with:
\begin{equation}
\label{4.1}  
\Delta_2(\delta,\theta_k) \;  = \;  \Delta_2(\delta)  \; \cos{(2 \theta_k)} \; ,
\end{equation}
where $ \Delta_2(\delta)$ monotonically increase with underdoping. In fact, Eq.~(\ref{4.1}) is consistent with our Eq.~(\ref{3.32}) and the doping dependence of the pseudogap
(see Fig.~\ref{fig-3}).  However, it should be stressed that many experiments report a gap that deviates from the simple relation in Eq.~(\ref{4.1})~\cite{Tanaka:2006,He:2009,Pushp:2009,Hashimoto:2010,Reber:2012,Sakai:2012,Yoshida:2012,Hashimoto:2012}. 
In fact, several experimental studies are consistent with the presence of two gaps, namely the pseudogap and a peculiar nodal gap. We believe that this nodal
gap is related to the dynamics of the nodal Fermi liquid discussed in Sect.~\ref{4-4}. Then, we see that the nodal gap has naturally a d-wave symmetry with
gapless nodes. On the other hand, the pseudogap being defined as the energy needed to break the real-space two-hole bound state, does not need to depend
on the geometry of the Fermi surface.
However, a full account on this subject lies beyond the aim of the present paper. We plan to present a detailed discussion of this point in a separate paper. \\
As concern the pseudogap temperature, in  Fig.~\ref{fig-8}, where we plot $\frac{T^*}{T_c}$ versus  $\frac{2 \Delta_2}{k_B T^*}$ for various 
cuprates~\cite{Kugler:2001}, we compare the experimental data with our relation   Eq.~(\ref{3.36}) which can be rewritten as:
\begin{equation}
\label{4.2}  
\frac{2 \Delta_2}{k_B \, T^*} \; = \;  4.0 \;  . 
\end{equation}
Indeed, we see that the data are consistent with   Eq.~(\ref{4.2}). We also plot in Fig.~\ref{fig-8} the d-wave BCS relation~\footnote{ The d-wave BCS critical temperature will be discussed in Sect.~\ref{5}. }:
\begin{equation}
\label{4.3}  
\frac{2 \Delta_2}{k_B \, T^*} \; \simeq \;  4.3 \;  . 
\end{equation}
Since Eq.~(\ref{4.3}) is quite close to our Eq.~(\ref{4.2}), the data are consistent with both relations. In our opinion, this unfortunate circumstance is the origin of confusion and several misinterpretations in the literature. \\
In conclusion, according to our previous discussion, the observed gap opening in the density of states (the pseudogap) marks the onset of the hole pairing at rather high 
temperatures $T \lesssim T^*$, but the critical temperature is set by the onset of phase coherence at lower temperatures $T \lesssim T_c = T_{B-K-T}$. The above point of view
is supported by several experiments. We believe that the most convincing evidence comes from Ref.~\cite{Hetel:2007}.  In fact, by using ultrathin films of Ca-substituted
$YBa_2Cu_3O_{7-\delta}$, the authors of Ref.~\cite{Hetel:2007} found a clear evidence of the Berezinskii-Kosterlitz-Thouless transition. In particular, they observe a remarkable scaling of the critical temperature with $n_s$ in accordance with Eqs.~(\ref{3.39}) and (\ref{3.41}). Further support on this picture comes from measurements of the high-frequency conductivity reported in Ref.~\cite{Corson:1999}. \\
To summarize, we may conclude that the key features of the underdoped side of the phase diagram are controlled by very strong hole pairing that is phase-disordered by  fluctuations~\cite{Emery:1995}.  In the following subsections, from a wealth of possible material, we have chosen to discuss some selected topics that we think well illustrate the physics of the underdoped region of hole-doped high-temperature superconductors.
\subsection{\normalsize{Real space bound state in magnetic fields}}
\label{4-1}
We are interested in the dynamics of hole pairs in presence of an external magnetic field perpendicular to the copper-oxide plane:
\begin{equation}
\label{4.4}  
\vec{H} \; = \; \; \nabla \times \vec{A} \; \; \;  , \;  \; \; \vec{A} \; = \; (A_1(\vec{r}),  A_2(\vec{r}), 0) \; .  
\end{equation}
In this case the Schr\"odinger equation Eq.~(\ref{3.11}) becomes:
\begin{eqnarray}
\label{4.5} 
  \left [  \frac{ 1}{2 m^*_h} \left (  - i \hbar \nabla_{ \vec{r}_1}  -  \frac{e}{c} \vec{A}( \vec{r}_1)  \right )^2  + 
  \frac{ 1}{2 m^*_h}  \left (  - i \hbar \nabla_{ \vec{r}_2}  -  \frac{e}{c} \,\vec{A}( \vec{r}_2)  \right )^2     \right ]  
  \Phi( \vec{r}_1 , \vec{r}_2 ) 
   \nonumber \\
   + \; V(\vec{r}_1 - \vec{r}_2 )  \;   \Phi( \vec{r}_1 , \vec{r}_2 ) \;  =   \;  E  \Phi (\vec{r}_1 , \vec{r}_2 )  \;   \; .  
\end{eqnarray}
It is useful to work in the symmetric gauge:
\begin{equation}
\label{4.6}  
 \vec{A} \; = \; - \,  \frac{1}{2} \;   \vec{r} \;   \times   \; \vec{H} \; ,  
\end{equation}
where the angular momentum is conserved. Using Eq.~(\ref{3.12}) we recast  Eq.~(\ref{4.5}) into:
\begin{eqnarray}
\label{4.7} 
  \left [  \frac{ 1}{4 m^*_h} \left (  - i \hbar \nabla_{ \vec{R}}  -  \frac{2e}{c} \vec{A}( \vec{R})  \right )^2  + 
  \frac{ 1}{ m^*_h}  \left (  - i \hbar \nabla_{ \vec{r}}  -  \frac{e}{2c} \,\vec{A}( \vec{r})  \right )^2     \right ]  
  \Phi( \vec{R} , \vec{r}) 
   \nonumber \\
   + \; V(\vec{r} )  \;  \Phi( \vec{R} , \vec{r} ) \;  =   \;  E  \Phi (\vec{R} , \vec{r})  \;   \; .  
\end{eqnarray}
Thus, we can write:
\begin{equation}
\label{4.8}  
\Phi (\vec{R} , \vec{r})  \; = \;  \Psi (\vec{R} ) \; \varphi( \vec{r})  \; ,  
\end{equation}
and 
\begin{equation}
\label{4.9}  
E \; = \;  E_{cond} \; - \; \Delta  \; ,
\end{equation}
such that:
\begin{equation}
\label{4.10} 
 \frac{ 1}{4 m^*_h} \left [  - i \hbar \nabla_{ \vec{R}}  -  \frac{2e}{c} \vec{A}( \vec{R})  \right ]^2   \Psi (\vec{R} )  
\;  =   \;  E_{cond} \;  \Psi (\vec{R})  \;  , 
\end{equation}
\begin{equation}
\label{4.11} 
\left [  \frac{ 1}{ m^*_h}  \left ( - i \hbar \nabla_{ \vec{r}}  -  \frac{e}{2c} \,\vec{A}( \vec{r})  \right )^2   + \; V(\vec{r} ) \right ]    \varphi( \vec{r})
\;  =   \; - \; \Delta  \;  \varphi( \vec{r}) \;  .
\end{equation}
The wavefunction $ \Psi (\vec{R} ) $ describes the condensate. In fact, $ \Psi (\vec{R} )$ is obtained from $\Phi (\vec{R} , \vec{r})$ after averaging over the internal degree
of freedom. Moreover, if we write:
\begin{equation}
\label{4.12}  
\Psi (\vec{R} ) \; = \; | \Psi (\vec{R})| \; \exp{ [i \Theta(\vec{R})]}  \; ,
\end{equation}
we see that in presence of an external magnetic field the condensate acquires a non-uniform phase. For an almost uniform condensate we have:
\begin{equation}
\label{4.13}  
 | \Psi (\vec{R})| \; =  \; constant \;  ,
\end{equation}
and $\Theta(\vec{R})$ is a slowly varying function. In this case we obtain:
\begin{equation}
\label{4.14}  
 E_{cond} \;   \simeq  \;  \frac{1}{2} \; (2   m^*_h) \;  \vec{v}_s^{\, 2}(\vec{R}) \;  ,
\end{equation}
where the supercurrent velocity is given by the well-known relation~\cite{Tinkham:1996}:
\begin{equation}
\label{4.15}  
 2  \,  m^*_h \;  \vec{v}_s (\vec{R}) \; = \; \hbar \;   \nabla_{ \vec{R}}  \; \Theta(\vec{R}) \; - \;  \frac{2e}{c} \; \vec{A}( \vec{R})  \; .
\end{equation}
Let us, now, consider Eq.~(\ref{4.11}) in polar coordinates:
\begin{equation}
\label{4.16}
  \left [  \frac{ - \hbar^2}{ m^*_h} \left (   \frac{\partial^2}{\partial r^2} +   \frac{1}{r}  \frac{\partial}{\partial r}  +  
   \frac{1}{r^2}  \frac{\partial^2}{\partial \theta^2}   \right )  + 
  \frac{ e^2 H^2}{16  m^*_h c^2} \; r^2  +  \frac{i e \hbar H}{2  m^*_h c} \; \frac{\partial}{\partial \theta}      +  V(r)    \right ]    \varphi(r,\theta)
  =   -  \Delta    \varphi(r,\theta)  \;  .    
\end{equation}
Again, after writing:
\begin{equation}
\label{4.17} 
\varphi( r , \theta  )  \; =  \;   \frac{\exp{( i m \theta)} }{\sqrt{2 \pi}}  \;    \varphi_m( r )         \; ,
\end{equation}
we have:
\begin{equation}
\label{4.18}
  \left [  \frac{ - \hbar^2}{ m^*_h} \left (   \frac{d^2}{d r^2}  +    \frac{1}{r}  \frac{d}{d r}  - 
   \frac{m^2}{r^2}    \right )  + 
  \frac{ e^2 H^2}{16  m^*_h c^2} \; r^2  +  V(r)  \right ]    \varphi_m(r)
   =    -  ( \Delta_m(H)   +   \frac{ e \hbar m H}{2  m^*_h c} ) \;  \varphi_m(r)
 \end{equation}
This last equation shows that, as expected, the external magnetic field lifts the degeneracy $m \rightarrow - m$.
\subsection{\normalsize{The upper critical magnetic field $H_{c2}$}}
\label{4-2}
We have already seen that for $H=0$ two holes give rise to a d-wave bound state. In presence of an external magnetic field the d-wave bound states are the solutions
of our pervious  Eq.~(\ref{4.18}) specialized to $m = \pm 2$:
\begin{equation}
\label{4.19}
  \left [  \frac{ - \hbar^2}{ m^*_h} \left (   \frac{d^2}{d r^2}  +    \frac{1}{r}  \frac{d}{d r}  - 
   \frac{4}{r^2}    \right )  + 
  \frac{ e^2 H^2}{16  m^*_h c^2} \; r^2  +  V(r)  \right ]    \varphi_{\pm 2}(r)
   =    -  ( \Delta_{\pm 2}(H)   \pm   \frac{ e \hbar  H}{  m^*_h c} ) \;  \varphi_{\pm 2}(r) \; .
 \end{equation}
We may, now, introduce the pair-breaking critical filed $H^{pb}_{c_2}$ such that:
\begin{equation}
\label{4.20} 
 \Delta_{\pm 2}(H^{pb}_{c_2},\delta)  \; =  \;  0      \; .
\end{equation}
To estimate the pair-breaking critical magnetic field we note that:
\begin{equation}
\label{4.21} 
 \Delta_{\pm 2}(0,\delta)  \; =  \;   \Delta_{ 2}(\delta)      \; ,
\end{equation}
where the pseudogap   $\Delta_{ 2}(\delta)$, already discussed in Sect.~\ref{3-1}, can be approximated quite accurately by:
\begin{equation}
\label{4.22} 
 \Delta_{ 2}(\delta)  \; \simeq  \;   \Delta_{ 2}(0)  \; \left [ 1 \; - (\frac{\delta}{\delta^*})^{1.5} \right ]    \; \; , \; \;  \Delta_{ 2}(0) \; \simeq \; 41.91 \; mev \; . 
 \end{equation}
Moreover, it is easy to see that the Zeeman term can be neglected. In fact, using Eq.~(\ref{2.19}) we find:
\begin{equation}
\label{4.23} 
  \frac{ e \hbar  H}{  m^*_h c}   \; \simeq  \;   2.14 \; 10^{-5} \;  ev \;  H(T)  \; , 
 \end{equation}
where $H(T)$ means that the strength of the magnetic field is measured in Tesla~\footnote{ $1 T = 10^4 G$. Even though we are using CGS units, it is customary
 in the literature to express the applied magnetic field in Tesla.}. Accordingly, we get from Eq.~(\ref{4.19}):
\begin{equation}
\label{4.24} 
 \Delta_{\pm 2}(H,\delta)  \; \simeq  \;      \Delta_{ 2}(\delta)     \;  - \;   \frac{ e^2 H^2}{16  m^*_h c^2}  \;  <r^2> \; .  
 \end{equation}
Since $<r^2> \simeq \xi_0^2$, we obtain:
\begin{equation}
\label{4.25} 
H^{pb}_{c_2}  \; \simeq  \;     \sqrt{   \frac{16  m^*_h c^2}{ e^2 \xi_0^2} \;  \Delta_{ 2}(\delta)  }  \; \; ,  
 \end{equation}
which leads to:
\begin{equation}
\label{4.26} 
H^{pb}_{c_2}  \; \simeq  \;   1.90 \; 10^3 \; T   \sqrt{ 1 \; - (\frac{\delta}{\delta^*})^{1.5}   }  \; \; .
 \end{equation}
\begin{figure}[t]
\begin{center}
\includegraphics[width=0.8\textwidth,clip]{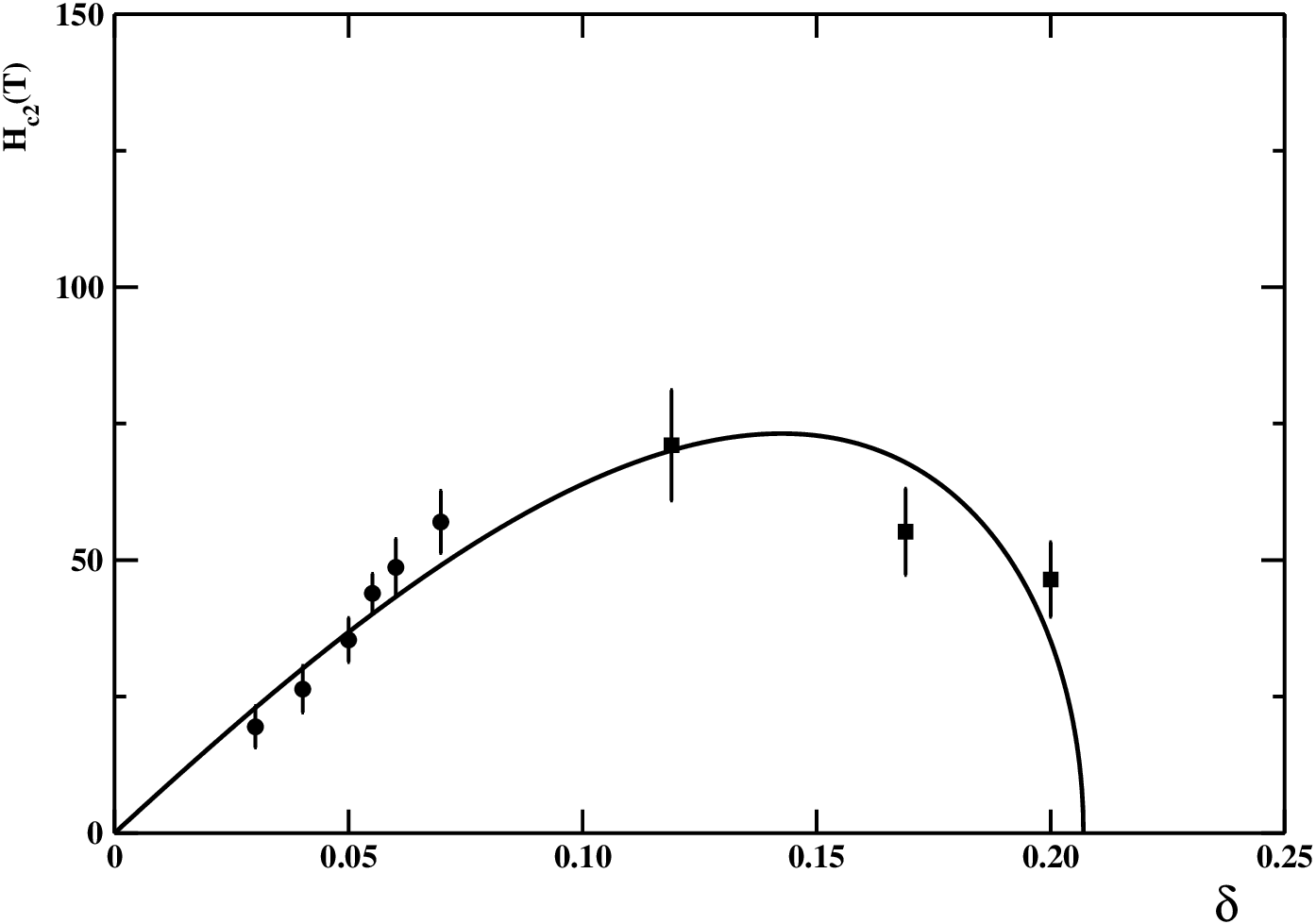}
\caption{\label{fig-9}  The upper critical field $H_{c2}$ versus the hole doping fraction $\delta$ according to Eq.~(\ref{4.31}) assuming $\kappa \simeq 1.3 \; 10^2$. 
 The data have been extracted from Fig.~18 of Ref.~\cite{Wang:2006} (full squares) and Fig.~4 of Ref.~\cite{Li:2007} (full circles).  }
\end{center}
\end{figure}
Equation (\ref{4.26}) shows that the pair-breaking critical field is very high. This means that the pseudogap is not affected by applied magnetic fields 
$ H \lesssim 10^2 \;T$
in accordance with several experimental observations. \\
In conventional superconductors the pair-breaking critical filed is of the same order of the depairing critical field. In our case, it turns out that the depairing critical field is
much smaller than the pair-breaking critical field. As a consequence, the upper critical field $H_{c_2}$ is determined by the depairing field. It is worthwhile to recall that in
type-II superconductors in the mixed state (Schubnikov phase) between the lower critical field $H_{c_1}$ and the upper critical field  $H_{c_2}$, the magnetic flux penetrates in a regular array of flux tubes, each carrying a quantum  flux:
\begin{equation}
\label{4.27} 
\phi_0  \; = \;   \frac{ 2 \pi \hbar c}{2 e} \; \simeq \; 2.07 \; 10^{-7}  \; G \; cm^2 \;  \; .
 \end{equation}
Within each unit cell of the array there is a vortex of supercurrent concentrating the flux toward the vortex center. To estimate the critical velocity $\vec{v}_c$ we use
Eqs.~(\ref{4.9}) and (\ref{4.14}) to get:
\begin{equation}
\label{4.28} 
E_{cond} \; \simeq   \; m^*_h \;  \vec{v}_c^{\, 2} \; \simeq  \Delta_{ 2}(H,\delta)    \; \simeq   \; \Delta_{ 2}(\delta) \; .    
 \end{equation}
Accordingly, the critical supercurrent is:
\begin{equation}
\label{4.29} 
\vec{J}_{c} \; = \; 2 \, e \, n_s \; \vec{v}_c \; \simeq   \;   e \,  \frac{\delta}{a_0^2}  \; \vec{v}_c \;  \; .    
 \end{equation}
From the  Maxwell equation $\nabla \times \vec{H} = \frac{4 \pi}{c} \vec{J}$ we easily get the following estimate:
\begin{equation}
\label{4.30} 
H_{c_2} \; \simeq \;  \frac{4 \pi}{c} \; |\vec{J}_c| \; \simeq \;    \frac{4 \pi}{c} \;  \frac{ e \delta}{a_0^2} \; 
 \sqrt{ \frac{\Delta_2(\delta)}{m^*_h}   }   \; .    
 \end{equation}
Since the magnetic field can be considered almost uniform over distance of order of the London penetration length $\lambda$, and considering that there are
$N_{vort} \simeq (\frac{\lambda}{\xi})^2 = \kappa^2$ vortices in a region of area $\lambda^2$, we obtain our final estimate of the upper critical field:
\begin{equation}
\label{4.31} 
H_{c_2}(\delta)  \; \simeq  \;  \kappa^2 \;    \frac{4 \pi}{c} \;  \frac{ e \delta}{a_0^2} \; 
 \sqrt{ \frac{\Delta_2(\delta)}{m^*_h}   }   \; .    
 \end{equation}
In Fig.~\ref{fig-9} we display our estimate of the upper critical field  Eq.~(\ref{4.31}) in the underdoped region $\delta < \delta^*$. 
We see that in the deep underdoped region the upper critical field decreases with underdoping in accordance with recent 
measurements~\cite{Wang:2006,Li:2007,Chang:2012}. In Fig.~\ref{fig-9} we also display the depairing critical field
as determined from the Nernst signal in  Refs.~\cite{Wang:2006,Li:2007}. It is known that the cuprate superconductors have a large Ginzburg-Landau parameter 
$\kappa \sim 10^2$ and are, therefore, extreme type-II superconductors. In fact, to compare quantitatively Eq.~(\ref{4.31}) with experimental data we choose 
$\kappa \simeq 130$.  It is gratifying to see that the doping dependence of our critical magnetic field as implied by Eq.~(\ref{4.31}) seems to follows quite closely the
experimental data.
\subsection{\normalsize{The vortex region}}
\label{4-3}
In the underdoped region of  the high-temperature cuprate superconductors there is convincing evidence for vortices at temperatures significantly above the critical
temperature~\cite{Wang:2006,Li:2007,Xu:2000,Capan:2002}. Indeed, our previous determination of the upper critical magnetic field shows that $H_{c_2}(\delta)$
is different from zero even for $\delta < \delta_{min} \simeq 0.05$, i.e. outside the superconducting dome in the phase diagram. To strengthen this point, following
Ref.~\cite{Li:2007b}, we define a critical vortex temperature $T_V(\delta)$ by the phenomenological relation:
\begin{equation}
\label{4.32} 
k_b \; T_V(\delta) \; \simeq \; 2.1 \; \mu_B \; H_{c_2}(\delta)    
 \end{equation}
where $\mu_B$ is the Bohr magneton. The meaning of $T_V(\delta)$ is that vortices are absent for $T \, > \, T_V(\delta)$. 
\begin{figure}[t]
\begin{center}
\includegraphics[width=0.8\textwidth,clip]{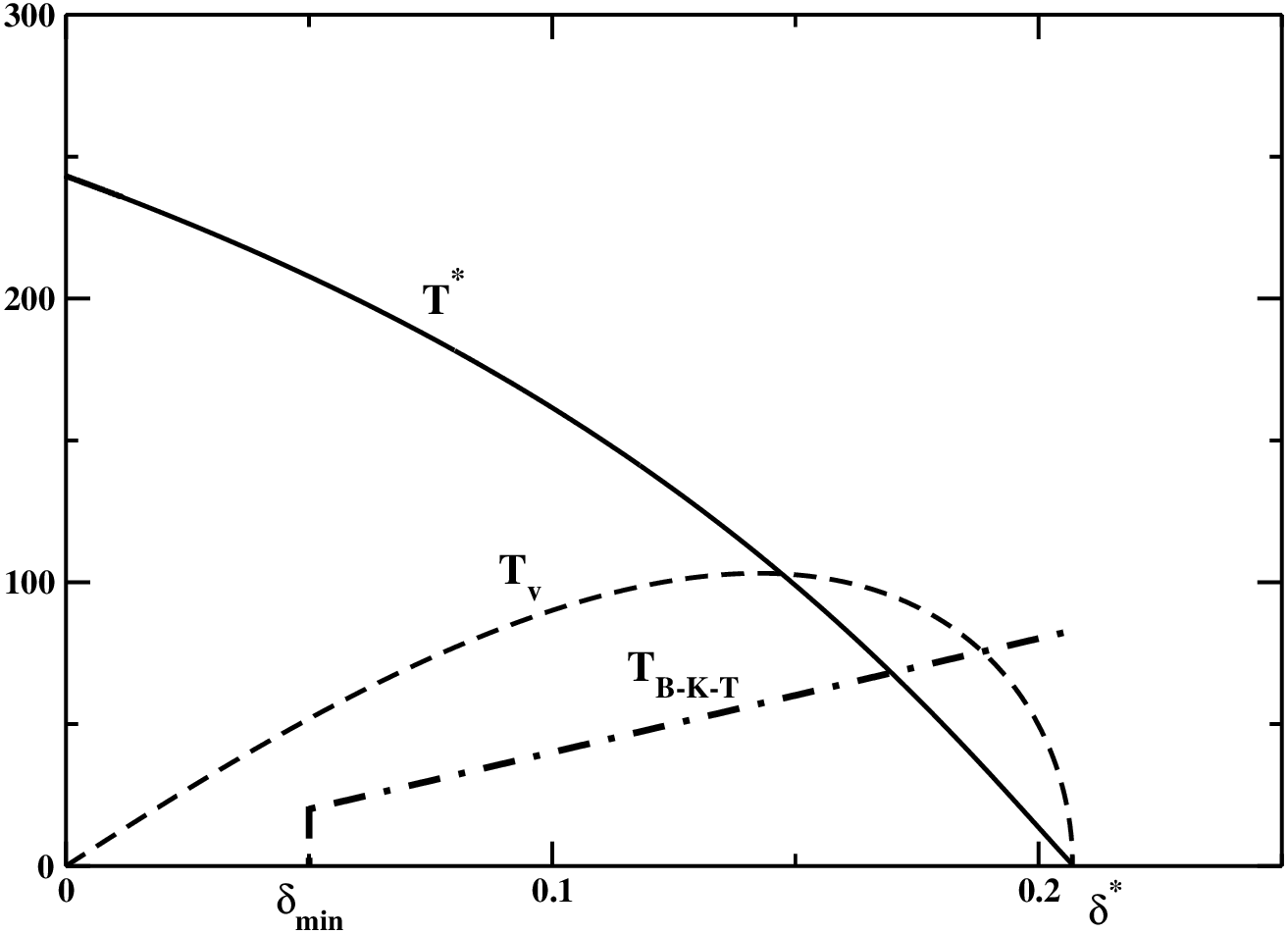}
\caption{\label{fig-10}  The pseudogap,  Berezinskii-Kosterlitz-Thouless, and vortex temperatures (in Kelvin) versus  $\delta $. The vortex region lies in
between $T_V$ and  $T_{B-K-T}$.}
\end{center}
\end{figure}
In Fig.~\ref{fig-10} we compare the vortex temperature to the critical Berezinskii-Kosterlitz-Thouless and  pseudogap temperatures. For temperatures 
$ T_{B-K-T}(\delta)  \, <  \, T \, <  \, T_V(\delta)$ (the so-called Nernst region) one should observe a large Nernst signal well above the superconducting critical temperature, in accordance with observations. \\
We would like to conclude this Section by addressing the puzzling observations of discrete core bound states in isolated vortices. 
In a classical paper~\cite{Caroli:1964}, it was predicted the existence of electronic states bound to the core of an isolated vortex at energies below the superconducting gap in the case of s-wave BCS superconductors.
Since the spacing of the discrete spectrum turns out to be very small, these bound states lead to a continuum which manifests itself as a zero-bias conductance peak in
scanning tunneling spectroscopy measurements. In high-temperature superconductors the scanning tunneling spectroscopy of isolated vortices revealed striking
electronic signatures which cannot be understood within the conventional BCS s-wave or d-wave theory~\footnote{ See Ref.~\cite{Fischer:2007}
 and references therein.}.
Indeed, the vortex cores give evidence of a pair of states with energy which scales with the superconducting gap and do not depend on the applied magnetic field.
If we denote with $E_{core}$ the energy of the vortex core states measured with respect to the zero-bias voltage, then there is a simple relation between the state energy
and the superconducting gap~\cite{Fischer:2007}: 
\begin{equation}
\label{4.33} 
E_{core}  \; \simeq \; \pm \; 0.3 \; \Delta_p  \; , 
 \end{equation}
where the slope were obtained by a linear fit to all data points, and $\Delta_p$ is the superconducting gap. We have already noted that the data are consistent with the
relation (see Fig.~\ref{fig-8}):
\begin{equation}
\label{4.34} 
\frac{ 2 \, \Delta_p}{k_B \, T^*}  \; \simeq  \; 4.3 \;   \; , 
 \end{equation}
so that, using  Eq.~(\ref{4.2}), we may rewrite Eq.~(\ref{4.33}) in terms of the pseudogap:
\begin{equation}
\label{4.35} 
E_{core}  \; \simeq \; \pm \; 0.32 \; \Delta_2  \; .
 \end{equation}
To interpret this last remarkable relation, we refer to the Schr\"odinger equation for the d-wave hole bound states in presence of an external magnetic field, Eq.~(\ref{4.18})
with $m = \pm 2$. We have shown that, neglecting the Zeeman splitting, $\Delta_{\pm 2}(H,\delta) \simeq \Delta_2(\delta)$. Thus, we infer that the pseudogap without the
external magnetic field fixes the zero-bias voltage in the core of an isolated vortex. Moreover, from Eq.~(\ref{4.18}) it follows that there are two states whose energies
measured with respect to the pseudogap are given by:
\begin{equation}
\label{4.36} 
E_{core}  \; = \; \pm \; \frac{e \hbar H_{core}}{m^*_h c}  \; ,
 \end{equation}
where $H_{core}$ is the magnetic field inside the vortex core. To estimate $H_{core}$, we note that the size of the isolated vortex can be determined by:
\begin{equation}
\label{4.37} 
\xi_V^2 \;  \simeq \; \frac{ \hbar^2}{\frac{m^*_h}{2} \Delta_2}  \; ,
 \end{equation}
where $\frac{m^*_h}{2}$ is the hole reduced mass. Thus, the magnetic field in the vortex core can be obtained from:
\begin{equation}
\label{4.38} 
\pi \; \xi_V^2 \;  H_{core} \;  \simeq \;\phi_0  \; ,
 \end{equation}
i.e.
\begin{equation}
\label{4.39} 
  H_{core} \;  \simeq \;  \frac{2 \pi \hbar c}{2 e} \;  \frac{1}{ \pi  \xi_V^2}  \;  .
 \end{equation}
Inserting  Eqs.~(\ref{4.39}) and Eq.~(\ref{4.37}) into  Eq.~(\ref{4.36}), we get:  
\begin{equation}
\label{4.40} 
E_{core}  \; \simeq \; \pm \; \frac{1}{2} \; \Delta_2  
\end{equation}
which is in satisfying agreement with  Eq.~(\ref{4.35}).
\subsection{\normalsize{Fermi arcs and quantum oscillations}}
\label{4-4}
\begin{figure}[t]
\begin{center}
\includegraphics[width=0.8\textwidth,clip]{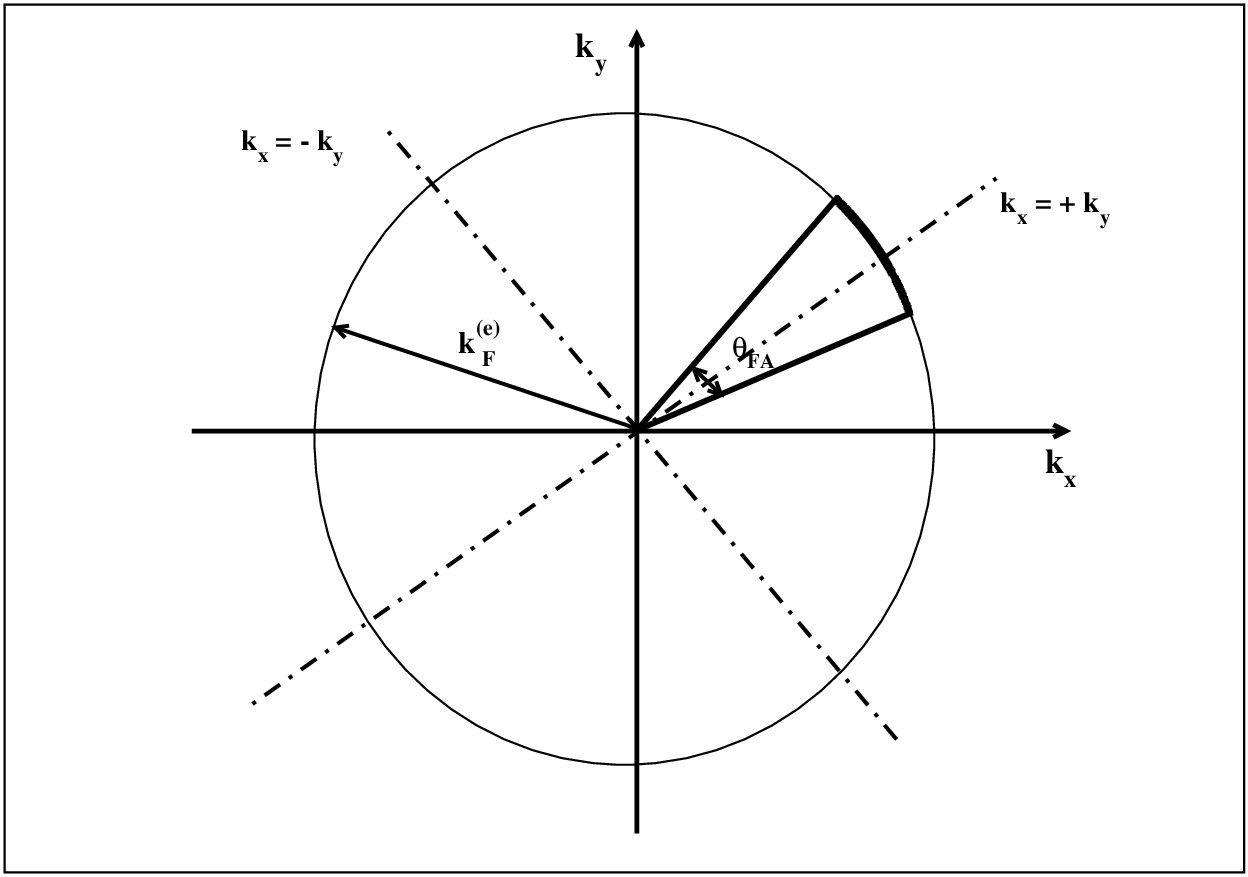}
\caption{\label{fig-11}  The Fermi sector and Fermi arc in the first quadrant of the Brillouin zone.}
\end{center}
\end{figure}

The normal-state properties in the underdoped region of high-temperature cuprates are highly anomalous. It is now well-established by many angle-resolved
photoemission studies~\footnote{ See Ref.~\cite{Damascelli:2003} and references therein.} that low-energy excitations are characterized by Fermi arcs,
namely truncate segments of a Fermi surface. The resistivity obeys $\rho \sim T$ over a wide range of temperatures in striking contrast to a Fermi-liquid
law $\rho \sim T^2$, expected for a conventional metals. Moreover, several recent
 studies~\cite{Sebastian:2012,Doiron:2007,Yelland:2008,Bangura:2008,Jaudet:2009,Bangura:2010}
report unambiguous identification of quantum oscillations in high magnetic fields. Interestingly enough, the measured low oscillation frequencies reveals a
Fermi surface made of small pockets. In fact, from the Luttinger's theorem and the Onsanger relation~\cite{Abrikosov:1972,Shoenberg:1984} between the
frequency and the cross-sectional area of the orbit, it turns out that the area of the pocket correspond to about $2 - 3 \;  \%$ of the first Brillouin zone area
in sharp contrast to that of overdoped cuprates where the frequency corresponds to a large hole Fermi surface~\cite{Bangura:2010}.  In addition, the authors
of Ref.~\cite{LeBoeuf:2007}  reported the observation of a negative Hall resistance in the magnetic-field -induced normal state, which reveals that these
pockets are electron-like rather than hole-like. More recently~\cite{Sebastian:2011}, experiments on the second harmonic quantum oscillations in underdoped
high-temperature cuprates lead to the conclusion that there exists only a single underlying quasi-two dimensional Fermi surface pocket. Moreover, it turns out
that the pocket is most likely associated with states near the nodal region of the Brillouin zone. This nodal pocket comprises quasielectron carrier to  explain both
the high field quantum oscillations with negative Hall and Seebeck~\cite{Laliberte:2011} effect. Finally, the recent studies of $YBa_2Cu_3O_{6.56}$ at very
high magnetic fields indicate that the specific heat exhibits the conventional temperature dependence and quantum oscillations expected for a Fermi liquid.
Moreover, the magnetic field dependence of the quasiparticle density of states follows the $\sqrt{H}$ behavior pointing to a d-wave superconducting gap.
Thus, the specific heat data demonstrate the surprising coexistence of the signature of a Fermi liquid and a d-wave superconducting gap over the entire magnetic
field range measured. \\
In the present Section we show that in our approach these features can be accounted for as a consequence of the pseudogap $\Delta_2(\delta)$. We have already
seen that, in the underdoped region $\delta < \delta^*$, the holes are bound into pairs with k-space wavefunction given by  Eq.~(\ref{3.32}). Even though  the pair
wavefunction vanishes along the nodal directions $k_x = \pm k_y$ ($\theta_k = \pm \frac{\pi}{4}$), there are not nodal low-lying hole excitations. This is due to the fact that
the pairing of the holes is in the real space and not in momentum space. On the other hand, due to the rotational symmetry, we may freely perform rigid rotations of
the pairs without spending energy. The rigid rotations of pairs is equivalent to hopping of electrons according to the hopping term in the Hamiltonian Eq.~(\ref{1.1}): 
\begin{equation}
\label{4.41}
\hat{H}_0^{(e)}  \; = \; -t \; \sum_{<i,j>,\sigma}   \left [  \hat{c}^{\dagger}_{i,\sigma}  \, \hat{c}_{j,\sigma}  \,  + \,   \hat{c}^{\dagger}_{j,\sigma}  \, \hat{c}_{i,\sigma} \right]  \; .
\end{equation}
We may diagonalize this Hamiltonian by writing:
\begin{equation}
\label{4.42} 
 \hat{\psi}_{e}(\vec{k},\sigma)  \; = \;  \frac{1}{\sqrt{M}}  \; \sum_{j}  \exp{(- i \vec{k} \cdot \vec{r}_j)}    \;  \hat{c}_{j,\sigma}   \; ,
 \end{equation}
to get:
\begin{equation}
\label{4.43} 
\hat{H}_0^{(e)}    \; = \;    \; \sum_{\vec{k},\sigma} \;  \varepsilon_{\vec{k}}^{(e)}   \;  \;  \hat{\psi}^{\dagger}_{e}(\vec{k},\sigma)  \;   \hat{\psi}_{e}(\vec{k},\sigma) \; ,
 \end{equation}
\begin{equation}
\label{4.44} 
\varepsilon_{\vec{k}}^{(e)}  \; = \;  -  \; 2 \, t   \; \left [ \; \cos{( k_x a_0}) \; +    \cos{( k_y a_0}) \;  \right ] \; . 
 \end{equation}
In the small-k limit we have:
\begin{equation}
\label{4.45} 
\hat{H}_0^{(e)}  \; = \;    \;  \sum_{\vec{k},\sigma} \;  \frac{\hbar^2 \, \vec{k}^2}{2 \, m^*_e}   \;  \;  \hat{\psi}^{\dagger}_{e}(\vec{k},\sigma)  \;   \hat{\psi}_{e}(\vec{k},\sigma) \; , 
 \end{equation}
where:
\begin{equation}
\label{4.46} 
m^*_e \; = \;  \frac{\hbar^2}{2  \, t  \, a_0^2} \; \simeq \; 2.17 \; m_e \;  . 
\end{equation}
Since there are $1 - \delta$ electrons per $Cu$ atoms, from the Hamiltonian Eq.~(\ref{4.45}) we may determine the electron Fermi energy:
\begin{equation}
\label{4.47} 
\varepsilon_{F}^{(e)}  \; = \;  \frac{\hbar^2 \, (\vec{k}_F^{(e)})^2}{2 \, m^*_e}  \;  \; \; , \; \; \;  a_0 \; k_F^{(e)} \; \simeq \;\sqrt{ 2 \pi (1 - \delta)}   \;  . 
\end{equation}
At first glance, one expects that the quasielectrons fill in momentum space the circle with radius $ k_F^{(e)}$ (the electron Fermi circle). However, one should keep in
mind that the hopping of electrons is possible thanks to the paired holes. Since in momentum space the wavefunction of a given pair is spread over a region around 
$k \sim \frac{1}{\xi_0}$, we see that the wavefunction of quasielectrons is likewise localized on a region in k-space around $\frac{1}{\xi_0}$. Thus, the quasielectrons
do not have the needed coherence to propagate over large distances with a well defined momentum. However, this argument does not apply along the nodal directions
where the momentum space hole-pair wavefunction vanishes. Therefore we are led to the conclusions that there are coherent quasielectrons that fill small circular sectors of the electron Fermi circle around the nodal directions  $k_x = \pm k_y$. Thanks to the rotational symmetry, we have four circular sectors with the same area. Since the number of coherent quasielectrons is determined by the doping fraction of holes (assuming that all the holes are paired), we obtain (see Fig.~\ref{fig-11}):
\begin{equation}
\label{4.48} 
\frac{\delta}{a_0^2} \;  \simeq \;  \frac{4 \times 2}{(2 \pi)^2 } \; \frac{1}{2} \;  (k_F^{(e)})^2 \; \theta_{FA} \; ,
\end{equation}
i.e.
\begin{equation}
\label{4.49} 
 \theta_{FA} \; \simeq \; \frac{\pi}{2} \; \frac{\delta}{1 - \delta}  \;  \; .
\end{equation}
We see, thus, that the Fermi surface is made by four Fermi arcs in agreement with the angle-resolved photoemission data. Moreover, the area of the Fermi sector with respect to the area of the Fermi circle in overdoped region turns out to be:
\begin{equation}
\label{4.50} 
\frac{{\cal{A}}_{FA}}{ {\cal{A}}_{overdoped}}\; \simeq \; \frac{1}{4} \; \frac{\delta}{1 + \delta}  \;  \; .
\end{equation}
For the typical value of hole doping fraction $\delta \simeq 0.1$, we infer from Eq.~(\ref{4.50}) that ${\cal{A}}_{FA}$ is about $ 2.3 \; \%$ of the first Brillouin zone area in the overdoped region, in satisfying agreement with quantum oscillation experimental results. Note that the effective quasielectron mass measured in quantum oscillation experiments  lies in the range $m^*_e \; = (2.0  -  3.0) \, m_e$, in agreement with our estimate  Eq.~(\ref{4.46}). \\
The drastic reduction of the available states in momentum space leads to a peculiar quasielectron conductivity. To see this, we may employ the simple Drude 
model~\cite{Ashcroft:1976}  for the frequency-dependent conductivity:
\begin{equation}
\label{4.51} 
\sigma(\omega) \; \simeq \; \frac{\sigma_0}{1 - i \omega \tau} \;  \; \; , \; \; \; \sigma_0 \; = \; \frac{n_e e^2 \tau}{m^*_e}  \;  \; ,
\end{equation}
where, according to Eq.~(\ref{4.48}), $n_e \simeq \frac{\delta}{a_0^2}$. Since the nodal quasielectrons suffer almost unidimensional scatterings, we get for the relaxation time the quite general estimate:
\begin{equation}
\label{4.52} 
\frac{\hbar}{\tau} \; \sim \;   \int_{ \varepsilon_{F}^{(e)} - \varepsilon}^{\varepsilon_{F}^{(e)}} d \varepsilon' \; \sim \; \varepsilon \; \sim k_B \; T  \; ,
\end{equation}
and Eq.~(\ref{4.51}) becomes:
\begin{equation}
\label{4.53} 
\sigma(\omega) \; \simeq \; \frac{\sigma_0}{1 - i \; \frac{\hbar \omega}{k_B T}} \;  \; .
\end{equation}
For $\hbar \omega \ll k_B T$ we obtain the conductivity:
\begin{equation}
\label{4.54} 
\sigma \; \simeq \; \sigma_0 \; \simeq \frac{n_e e^2}{m^*_e}  \; \frac{\hbar}{ k_B T} \; \simeq    \frac{e^2 \; \delta}{a_0^2 \; m^*_e}  \; \frac{\hbar}{ k_B T} \; \;  .
\end{equation}
On the other hand, if $\hbar \omega \gg k_B T$ we get:
\begin{equation}
\label{4.55} 
\sigma(\omega) \; \simeq \; \frac{\sigma_0}{- i \omega \tau} \;   \simeq i \;  \frac{n_e e^2}{m^*_e \; \omega }   \; \simeq   \; 
i \;  \frac{e^2 \; \delta}{a_0^2 \; m^*_e}  \; \frac{1}{ \omega} \; \;  .
\end{equation}
These relations are in qualitative agreement with several observations. In particular  Eq.~(\ref{4.54}) should be valid up to temperatures of the order of the pseudogap
temperature $T^*$. \\
Let us conclude this Section by discussing the peculiar dependence of the specific heat on the applied magnetic field reported in Ref.~\cite{Riggs:2011}. 
Firstly, we evaluate the Sommerfeld coefficient for $H=0$. A standard calculation gives:
\begin{equation}
\label{4.56} 
\gamma_s  \; \simeq \; \frac{\delta}{ 1  - \delta} \; \frac{\pi}{3} \;   \frac{m^*_e a_0^2}{ \hbar^2 }   \; N_A \; k_B^2
\end{equation}
where $N_A$ is the Avogadro's number.  The authors of Ref.~\cite{Riggs:2011} found that at very high magnetic fields the specific heat in an underdoped high-temperature superconductors exhibits both the conventional temperature dependence and quantum oscillations expected for a Fermi liquid. The oscillatory component of the specific heat is given by the Lifshitz-Kosevich formula~\cite{Shoenberg:1984}. On the other hand, for the Sommerfeld coefficient they report: 
\begin{equation}
\label{4.57} 
\gamma_s(H) \; = \;  \gamma_s(H=0) \; + A_c \; \sqrt{H} \; ,
\end{equation}
with
\begin{equation}
\label{4.58} 
  \gamma_s(H=0) \; = \; 1.85 \; \pm \; 0.06 \; mJ \;  mol^{-1} K^{-2} \; \; , \; \;  A_c \; \simeq  \;  0.47  \; mJ \; mol^{-1} \;  K^{-2} \; T^{-\frac{1}{2}} \; . 
\end{equation}
To compare our Eq.~(\ref{4.56})  with Eq.~(\ref{4.58}) one should keep in mind that there are two $CuO_2$ planes per formula unit in $YBCO$. Moreover, the planar
$CuO_2$ densities in high-temperature cuprates are difficult to estimate. Assuming $\delta \simeq 0.1$ and using the effective electron mass Eq.~(\ref{4.46}), we find
(in MKS units):
\begin{equation}
\label{4.59} 
  \gamma_s \; \simeq \; 0.76  \; mJ \;  mol^{-1} K^{-2} \; \; , 
\end{equation}
which differs from measured value in Eq.~(\ref{4.58}) by about a factor of $2$.  As far as  field-induced Sommerfeld coefficient is concerned, the behavior $\gamma_s(H) \sim \sqrt{H}$ is interpreted as evidence of the presence of a d-wave superconducting gap. In fact, it was pointed out~\cite{Volovik:1993} that the increase of the Sommerfeld coefficient with  the external magnetic field is due to the line nodes of the superconducting gap. The evidence of quantum oscillations and field-induced Sommerfeld coefficient in the electronic specific heat points to the surprising coexistence of a Fermi liquid and a d-wave superconductor. In our approach, we now show that the quasielectron nodal Fermi liquid leads
naturally to the observed Sommerfeld coefficient $\gamma_s(H) \sim \sqrt{H}$. In the mixed state, according to Ref.~\cite{Volovik:1993} the energies of the low-lying
excitations circulating around a vortex are shifted by Doppler effect. Since it results that $k_F^{(e)} \xi_0 \gg 1$, we may employ the semiclassical approach which considers the momentum and position of quasiparticles as commuting variables. The effects of the supercurrent circulating around a vortex is accounted for by a Doppler shift of  the quasiparticle energy. Thus, for the quasielectrons we have:
\begin{equation}
\label{4.60} 
  \varepsilon(\vec{k}, \vec{r})   \; \simeq  \;    \frac{\hbar^2 \vec{k}^2}{2 \; m^*_e}  \; + \;  \hbar \; \vec{k} \cdot \vec{v}_s(\vec{r})         \; ,
 \end{equation}
where $\vec{v}_s(\vec{r}) $ is the supervelocity that, according to Eq.~(\ref{4.15}), is given by:
\begin{equation}
\label{4.61} 
 \vec{v}_s(\vec{r})  \; \simeq \frac{\hbar}{4 \; m^*_h} \; \frac{\hat{\theta}_r}{r}    \;   \; .
 \end{equation}
The Doppler shift  Eq.~(\ref{4.60}) has a sizable effect on the density of state. To see this, we need to evaluate the density of state at the Fermi surface
averaged over the vortex:
\begin{equation}
\label{4.62} 
{\cal{N}}(0) \; \equiv \; 2 \; \int \; \frac{d \vec{k}}{(2 \pi)^2} \;  \int \; d\vec{r} \;  \delta \left [  \varepsilon(\vec{k}, \vec{r})  -   \varepsilon_F^{(e)} \right ] \; 
\frac{1}{\pi R^2}  \; ,
 \end{equation}
where 
\begin{equation}
\label{4.63} 
R \; \simeq   \xi_V \; \sqrt{ \frac{H_{c_2}}{H} }
 \end{equation}
is the average distance between vortices.  We obtain:
\begin{equation}
\label{4.64} 
{\cal{N}}(0)  \simeq  \frac{2}{4 \pi^2}   \int_{FA} d\theta_k  \int_{0}^{\infty} dk \; k   \int d\theta_r  \int_{\xi_0}^{R}  dr 
 \frac{r}{\pi R^2} \;  \delta \left [ \frac{\hbar^2 k^2}{2  m^*_e}  -   \frac{\hbar^2 k \sin{(\theta_r+\theta_k)} }{4  m^*_e r}  - \varepsilon_F^{(e)}  \right ] 
 \end{equation}
where the angular integration on $\theta_k$ is performed over the four Fermi arcs ($FA$). The integration over $k$ can be done easily. One obtains:
\begin{equation}
\label{4.65} 
{\cal{N}}(0)  \; \simeq  \; \frac{1}{4 \pi^2}  \; \frac{(m^*_e)^2}{ \hbar^2 m^*_h} \;  \int_{FA} d\theta_k  \int d\theta_r  \int_{\xi_0}^{R}  
 \frac{dr}{\pi R^2} \;    \frac{\sin{(\theta_r+\theta_k)} }{ k_F^{(e)} }  \; .
 \end{equation}
Finally the angular integrations give:
\begin{equation}
\label{4.66} 
{\cal{N}}(0)  \; \simeq  \; \frac{1}{4 \pi^2}  \; \frac{(m^*_e)^2}{ \hbar^2 m^*_h} \;  
 \frac{2 \; \sqrt{2} \; \theta_{FA}}{\pi \;  k_F^{(e)} \;  R}  \; ,  
 \end{equation}
or, using Eqs.~(\ref{4.63})  and (\ref{4.49}):
\begin{equation}
\label{4.67} 
{\cal{N}}(0)  \; \simeq  \; \frac{\sqrt{2}}{4 \pi^2}  \;  \frac{\delta}{1 - \delta} \; \frac{(m^*_e)^2}{ \hbar^2 m^*_h} \;  
 \frac{1}{ k_F^{(e)} \;  R}  \; \sqrt{ \frac{H}{H_{c_2}} } \; \; .
 \end{equation}
Since the high-temperature superconductors are extreme type-II superconductors, one should average the density of state over the vortex distribution. However,
from Eq.~(\ref{4.67}) it is evident that the quasiparticle energy Doppler shift induced by vortices  leads to the expected field-induced Sommerfeld coefficient  
$\gamma_s(H) \sim \sqrt{H}$.
\section{The physics of the over and optimal doped regions}
\label{5}
In the overdoped region we have already stressed that the normal state properties can be described reasonably well by the Fermi liquid picture, although still with some
electronic correlations, with a hole density corresponding to $ 1 + \delta$ holes per $Cu$ atom. Indeed, as long as $\delta > \delta^*$,
the effective two-body potential  Eq.~(\ref{2.7}) can be dealt with as a small perturbation. Therefore, the dynamics of the copper-oxide layers is governed
essentially  by our effective Hamiltonian $\hat{H}_0$, Eq.~(\ref{2.18}). Note that the correlation effects due to the on-site repulsion $U$ are built into $\hat{H}_0$ 
by means of the effective hole mass $m^*_h$. It is interesting to note that:
\begin{equation}
\label{5.1} 
 \frac{m^*_h}{m^*_e}  \;  =  \;  \frac{U}{4 \; t}   \;  \simeq  \; 2.5 \; \;  .
 \end{equation}
In fact, several quantum oscillation experiments give consistently an effective quasiparticle mass of carrier in the overdoped region which is approximately twice the
effective quasiparticle mass of carriers in the underdoped region. \\
It is interesting to compare the low-temperature linear term in the electronic specific heat of the quasiholes with the one of the nodal quasielectrons evaluated
in Sect.~\ref{4-4}. Since the quasiholes behave like a two-dimensional normal Fermi liquid, we obtain instead of  Eq.~(\ref{4.56}):
\begin{equation}
\label{5.2} 
\gamma_s  \; \simeq \;  \frac{\pi}{3} \;   \frac{m^*_h a_0^2}{ \hbar^2 }   \; N_A \; k_B^2 \; .
\end{equation}
In fact, it is easy to check that the Sommerfeld coefficient given by  Eq.~(\ref{5.2}) exceeds by about an order of magnitude the quasielectron Sommerfeld coefficient.
Using our hole effective mass we obtain (MKS units):
\begin{equation}
\label{5.3} 
\gamma_s  \; \simeq \;  8.53  \; mJ \;  mol^{-1} K^{-2} \; \; .
\end{equation}
Our determination of $\gamma_s$  compares quite well with the observed value:
\begin{equation}
\label{5.4} 
\gamma_s  \; \simeq \;  7 \; \pm \; 1  \; mJ \;  mol^{-1} K^{-2} 
\end{equation}
as obtained from direct measurements in polycrystalline $Tl_2Ba_2CuO_{6+\delta}$~\cite{Wade:1994}, independently on $\delta$, and also from indirect
determination through the effective hole mass $m^*_h = (5.2 \pm 0.4) m_e$ in Ref.~\cite{Bangura:2010}. \\
Concerning the superconducting phase in the overdoped region, we already argued that this can be described within the conventional BCS framework. In  fact, there
is a large body of evidence that the weak-coupling d-wave BCS approach accounts for many of the low-energy and low-temperature properties of overdoped copper
oxides.  Here, we restrict ourself to the discussion of a clear signature of d-wave BCS pairing in momentum space, namely the ratio of the BCS gap to the critical
temperature anticipated in Eq.~(\ref{3.61}):
\begin{equation}
\label{5.5}  
\frac{\Delta_{BCS}} {k_B T_c} \; \simeq \; \frac{ 2 \;  \pi}{\sqrt{e}}  e^{ - \gamma}  \; \simeq \; 2.14 \;  \; .
\end{equation}
To derive Eq.~(\ref{5.5}), we observe that the gap equation  Eq.~(\ref{3.47}) at finite temperature reads:
\begin{equation}
\label{5.6}  
\Delta(\vec{k}) \;  = \; - \frac{1}{2} \;  \sum_{\vec{k}'} \; \; \frac{ V(\vec{k} - \vec{k}')  \Delta(\vec{k}')} { \sqrt{ \xi^2_{\vec{k}'} + |  \Delta(\vec{k}')|^2 } }  
\tanh{ \left [ \frac{\sqrt{ \xi^2_{\vec{k}'} + |  \Delta(\vec{k}')|^2 }}{2 \; k_B \; T} \right ] } \; .
\end{equation}
Thus, instead of Eq.~(\ref{3.56}) we have:
\begin{equation}
\label{5.7}  
1 \;  \simeq \; -   V_2  \; \int \frac{d\vec{k}'}{(2 \pi)^2}  \; 
 \; \frac{  [\cos{(2 \theta_{k'})}]^2 } { \sqrt{ \xi^2_{\vec{k}'} + [  \Delta_{BCS} \cos{(2 \theta_{k'})}]^2 } }  
 \tanh{ \left [ \frac{\sqrt{ \xi^2_{\vec{k}'} + [  \Delta_{BCS} \cos{(2 \theta_{k'})}]^2 }}{2 \; k_B \; T} \right ] } \; .
\end{equation}
Since the critical temperature is defined by  $\Delta_{BCS}(T_c)=0$, we obtain:
\begin{equation}
\label{5.8}  
1 \;  \simeq \; -   \frac{V_2 m^*_h}{\hbar^2}  \; \int_{-\varepsilon_c}^{+\varepsilon_c}  \frac{d \xi}{(2 \pi)^2}  \;  \int_{0}^{2 \pi} d\theta
 \; \frac{  [\cos{(2 \theta)}]^2 } {  | \xi | }  \;
  \tanh{ \left [ \frac{ | \xi | }{2 \; k_B \; T_c} \right ] }  \; ,
\end{equation}
i.e.
\begin{equation}
\label{5.9}  
1 \;  \simeq \; -   \frac{V_2 \;  m^*_h}{2  \pi \; \hbar^2}  \; \int_{0}^{\varepsilon_c}   \; \frac{d \xi}{\xi}    \; 
  \tanh{ \left [ \frac{ | \xi | }{2 \; k_B \; T_c} \right ] }  \; .
\end{equation}
Upon using Eq.~(\ref{3.60}) we may rewrite  Eq.~(\ref{5.9}) as:
\begin{equation}
\label{5.10}  
1 \;  \simeq \; \lambda_2 \;  \int_{0}^{ \frac{\varepsilon_c}{2k_BT_c}     }   \; \frac{d z }{z}    \;   \tanh{ [z]}  \; 
\simeq \; \lambda_2 \; \left [ \ln{(\frac{\varepsilon_c}{2k_BT_c} )} \; - \; \ln{(\frac{\pi}{4})} \; + 
\; \gamma \right ] \; ,
\end{equation}
which leads to:
\begin{equation}
\label{5.11}  
k_B \; T_c  \;  \simeq \;   \frac{\varepsilon_c}{2}   \;   \exp{ ( - \ln{\frac{\pi}{4}} + \gamma)}   \;  \exp{ [ -  \frac{1}{\lambda_2}]}  \; .
\end{equation}
Combining   Eq.~(\ref{5.11}) with   Eq.~(\ref{3.59}) one obtains  Eq.~(\ref{5.5}).
\begin{figure}[t]
\begin{center}
\includegraphics[width=0.8\textwidth,clip]{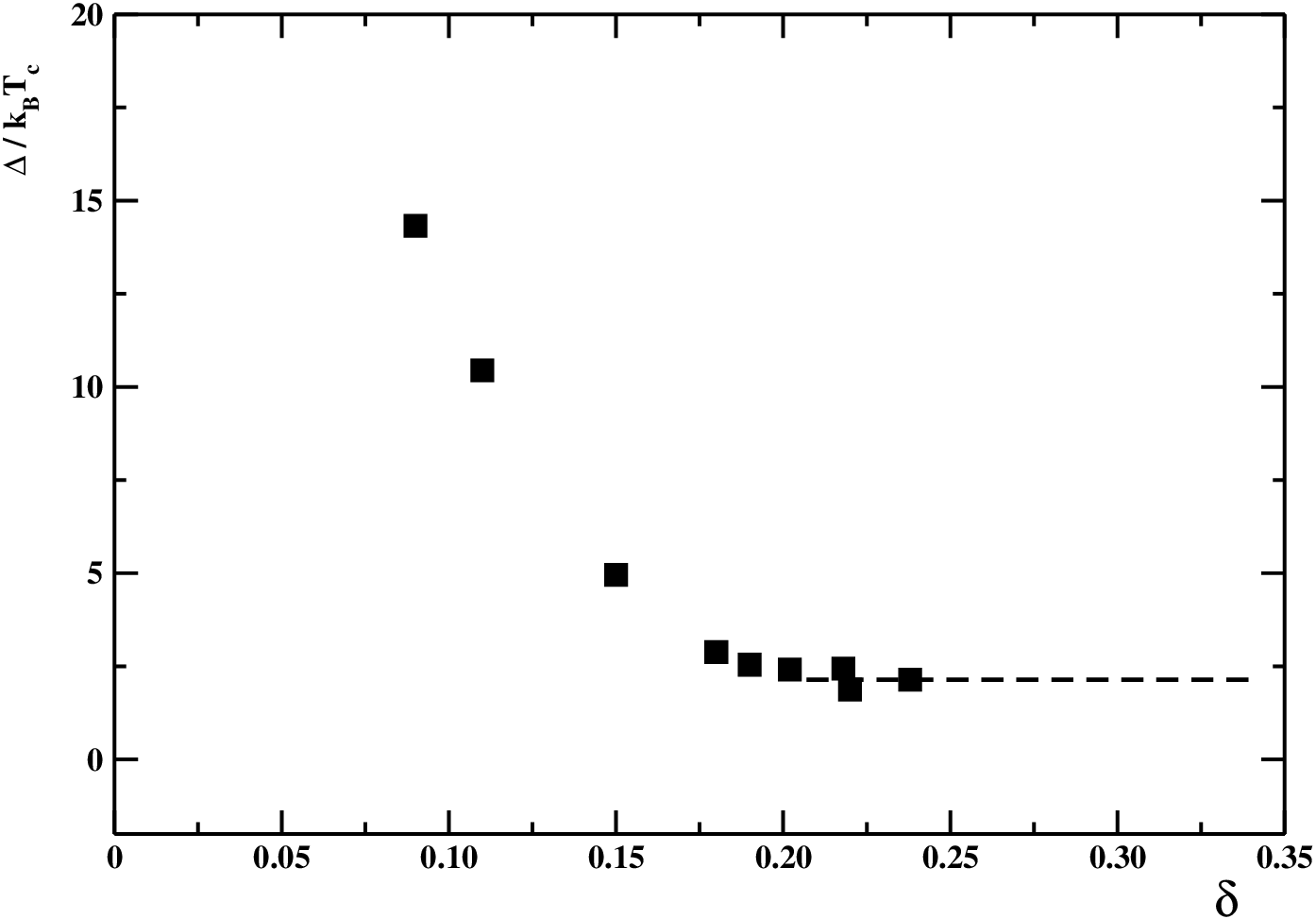}
\caption{\label{fig-12}  Doping dependence of the ratio of superconducting gap $\Delta$ over the critical temperature $T_c$.
The data has been extracted from Fig.~4 of Ref.~\cite{Wang:2007}. The dashed line is the weak coupling d-wave BCS value
$\frac{\Delta}{k_B T_c} \simeq 2.14$. }
\end{center}
\end{figure}
In Figure~\ref{fig-12} we display the doping dependence of the ratio Eq.~(\ref{5.5}) using the data reported in Fig.~4 of Ref.~\cite{Wang:2007}. 
It is evident that in the overdoped region $\delta \gtrsim 0.20$ the gap values approach closely onto the theoretical prediction of the weak-coupling
d-wave BCS superconductivity. \\
Let us now discuss the optimal doped region $\delta \sim \delta_{opt}$. According to our picture, the low-doping region is characterized by a large pseudogap
$\Delta_2(\delta)$ which decreases rapidly with increasing doping, while the superconductivity is driven by the finite-temperature  Berezinskii-Kosterlitz-Thouless 
transition with a critical temperature that increases with doping. On the other hand, in the overdoped region the superconducting transition is well described
by the conventional weak-coupling d-wave BCS pairing in momentum space. The decrease of the critical temperature with increasing doping level originates
from the underlying reduction in the pairing strength due to the disturbing of the antiferromagnetic correlations with overdoping. From our Fig.~\ref{fig-7} we
see that the two critical temperatures meet in the optimal doped region $\delta \sim \delta_{opt}$, explaining in a natural way why the critical superconducting temperature reaches here the maximum value. Since the   Berezinskii-Kosterlitz-Thouless temperature is comparable to the BCS critical temperature, this makes fluctuations much more important.
Even more,  Fig.~\ref{fig-7} suggests that also the pseudogap temperature $T^*$ becomes comparable to both $T_{B-K-T}$ and $T_{BCS}$. We conclude, thus, that
in this region of the phase diagram there are competing phases and it is quite difficult to precisely pin down the underlying physics.  On the other hand, as our 
Fig.~\ref{fig-7} is suggesting, there is a large body of evidence that pseudogap energy scale does not extend into the heavily overdoped region of the
phase diagram, but rather collapse at a well defined critical concentration around $\delta^* \simeq 0.20$. We would like to stress that the presence of an expanded
phase-fluctuation regime in the optimal doped region may explain why certain spectroscopic probes, like angle-resolved photoemissions, tend to advocate scenarios
in which the pseudogap temperature tracks the superconducting dome rather than vanishing inside it. The scenario advanced in this paper is supported by the
results presented in Ref.~\cite{Guyard:2008}.  By means of a systematic electronic Raman scattering study of a mercury-based single layer cuprate as a function of doping level, the authors of Ref.~\cite{Guyard:2008} revealed the existence of a breakpoint close to optimal doping, below which the antinodal gap (the pseudogap) is gradually disconnected from superconductivity. The nature of both the superconducting and normal state is distinctly different on each site of this breakpoint. This is consistent also with electronic specific heat measurements. In fact, it turns out that~\cite{Loram:1997,Loram:2001} in optimally and overdoped samples the Sommerfeld
coefficient is almost constant for $T > T_c$, in accord with our Eq.~(\ref{5.2}), while for underdoped samples there is considerable drop in the Sommerfeld
coefficient for $T < T^*$ (see our Eq.~(\ref{4.56})). Moreover, there is evidence for carrier density which decreases smoothly from $\frac{1+\delta}{a_0^2}$ in
the overdoped phase to $\frac{\delta}{a_0^2}$  in the underdoped phase. In fact, in Refs.~\cite{Daou:2009,Daou:2009b} it is shown that it exists a critical
point $\delta^*$ where the pseudogap temperature $T^*$ goes to zero and  which lies in the superconducting dome. Just below   $\delta^*$  there is a crossover from a metal with large hole-like Fermi surface, consistent with $n_h \simeq \frac{1+\delta}{a_0^2}$, to a metal with low density of charge carrier, which from Hall and Seebeck effect measurements  are quasielectrons, with $n_e \simeq \frac{\delta}{a_0^2}$. \\
Before concluding this Section, we would like to remark that the optimal doped region of hole-doped cuprates is the most difficult to precisely characterize.
In fact, according to our current theoretical understanding, in this region there is competition between the BCS gap and the pseudogap, and the presence
of sizable phase fluctuations strongly limit our mean-field treatment. Nevertheless, it is worthwhile to disciss the dependentce of the maximum critical temperature 
$T_c^{max}$ on the microscopic parameters of the model. As already discussed we have: 
\begin{equation}
\label{5.12}  
T_c^{max}  \;  \simeq \;  T_{BCS}(\delta_{opt} \simeq \delta^*)  \; .
\end{equation}
Now, from Eqs.~(\ref{2.5}),  (\ref{2.9}),  (\ref{2.17}),  (\ref{3.46}) and  (\ref{3.60}) we obtain:
\begin{equation}
\label{5.13}  
T_c^{max}  \;  \sim \;   \frac{t^2}{U}   \;  \exp{ [ -  \frac{1}{(1 - \delta^*/\delta_c)^2}]} 
\end{equation}
where we recall that $\delta^*$ is the hole doping fraction where the pseudogap $\Delta_2$ vanishes. Apparently Eq.~(\ref{5.13}) suggests that one can increase
$T_c^{max}$  by increasing $\frac{t^2}{U}$ and/or decreasing $\delta^*$. However, it turns out that the increase of   $\frac{t^2}{U}$ leads also to an increase of 
$\delta^*$ and vice versa, such that the two effects compensate in  Eq.~(\ref{5.13}). This explains why the optimal hole doping fraction is very close to $\delta^*$
and attains the universal value:
\begin{equation}
\label{5.14}  
\delta_{opt}  \;  \simeq \;  \delta^* \; \simeq \; 0.20  
\end{equation}
in the whole family of cuprates.
\section{The electron doped cuprates}
\label{6}
\begin{figure}[t]
\begin{center}
\includegraphics[width=0.8\textwidth,clip]{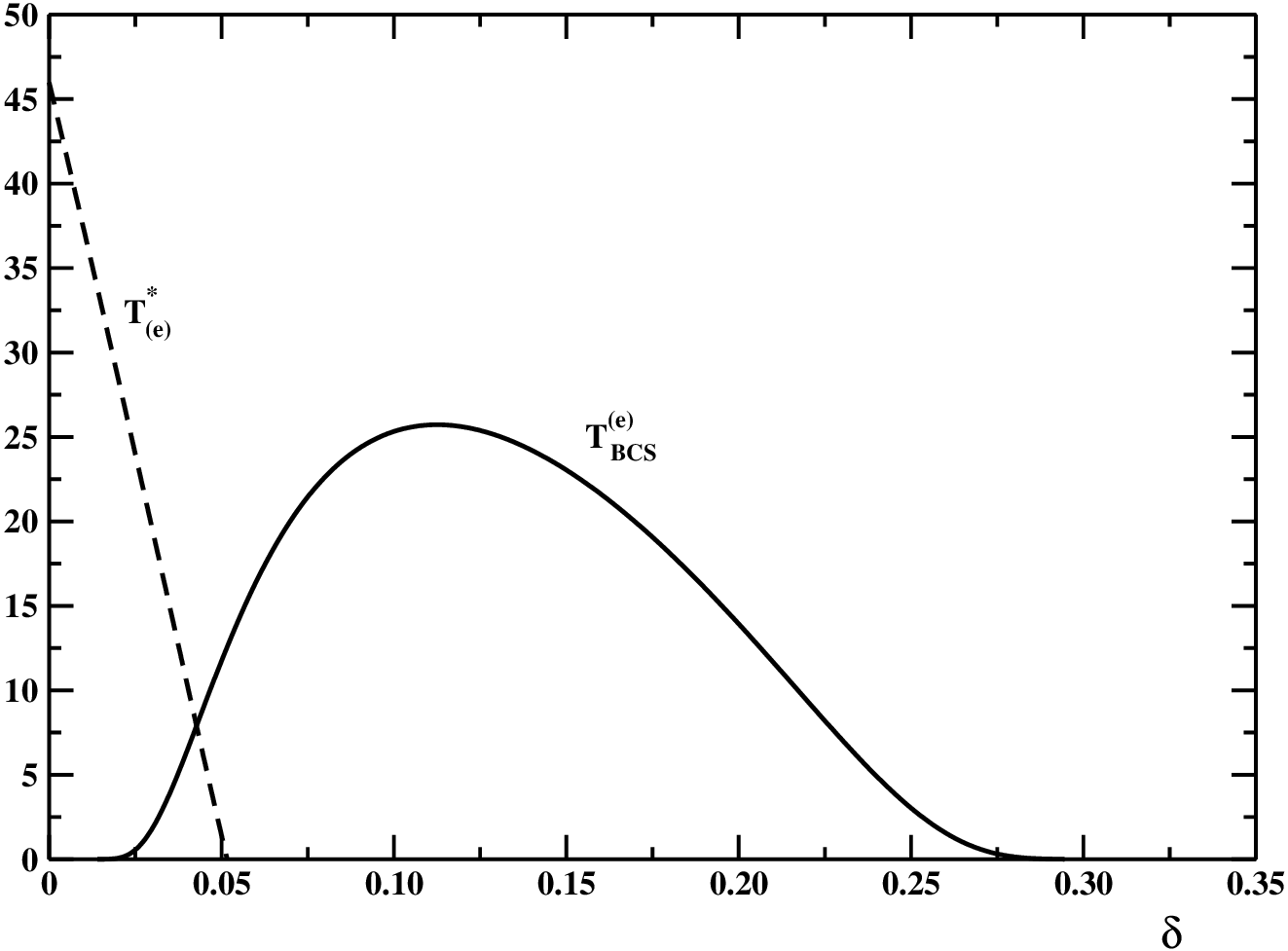}
\caption{\label{fig-13}  The pseudogap and  BCS critical  temperatures (in Kelvin) versus the doping fraction
for the electron-doped cuprates. }
\end{center}
\end{figure}

The vast majority of high-temperature superconductors are hole-doped. Nevertheless, in this Section we will touch upon high-temperature electron-doped
cuprates (see Ref.~\cite{Armitage:2010}  for a review). The most evident difference between electron and hole doped cuprates is that the antiferromagnetic
phase at $\delta=0$ is much more robust in the electron-doped material and it persists to much higher doping levels. If we adopt the effective single-band Hubbard
model of Sect.~\ref{1} for the microscopic description of the $CuO_2$ layers, then we see that the electrons injected into the copper-oxide plane would behave
like the holes in the overdoped region of the hole-doped cuprates. However, due to the stronger antiferromagnetic correlations in the ground state for $\delta > 0$,
the electron carrier density is:
\begin{equation}
\label{6.1}  
n_e  \;  \simeq \;   \frac{\delta}{a_0^2}   \;    \; ,
\end{equation}
since one electron of the $(1+\delta)$ electrons per Cu atom is frozen in the half-filled Hubbard band antiferromagnetic configuration. In our approach the effective Hamiltonian for the injected electrons should be given by   Eqs.~(\ref{2.18})   and (\ref{2.6}). In particular, the electron effective mass agrees with   Eq.~(\ref{2.17}): 
\begin{equation}
\label{6.2} 
m^*_e \; \simeq \;  \frac{\hbar^2}{8  \, \frac{t^2}{U} a_0^2}  \; \simeq \; 5.41 \; m_e  
\end{equation}
with the previously stated values of the model parameters. However, due to the increased role played by the fermion correlations in the antiferromagnetic background,
we expect that the effective two-body potential in the interaction Hamiltonian Eq.~(\ref{2.6}) is reduced to some extent. To take into account this effect, we shall assume
that the effective potential $V(\vec{r}_1- \vec{r}_2)$ has the same form as in Eq.~(\ref{2.7}) but with a somewhat reduced interaction length:
\begin{equation}
\label{6.3} 
 r_0^{(e)}(\delta) \; = \;  4.15 \;  a_0 \; \left ( 1 \; - \; \frac{\delta}{\delta_c} \right )^{\frac{1}{2}} \; \; \; , \; \; \delta_c \; \simeq \; 0.35  \; .
 \end{equation}
We now show  that this small variation in the range of the effective two-body interaction potential leads to a dramatic change in the phase diagram of the
electron-doped cuprates. In fact, by solving the real-space d-wave bound state Schr\"odinger equation Eq.~(\ref{3.11}) we find a drastically reduced pseudogap 
(see Fig.~\ref{fig-13}):
\begin{equation}
\label{6.4} 
 T^*_{(e)}(\delta=0) \; = \;  \frac{ \Delta_2^{(e)}(\delta=0)}{ 2 \; k_B} \; \simeq \; 46 \; K  \; \; \; , \; \; \; \delta^* \; \simeq \; 0.05 \; \; .
 \end{equation}
This means that, according to the discussion in Sect.~\ref{3-2}, the pseudogap superconducting phase of the underdoped region of hole-doped cuprates is absent.
Therefore, the superconducting transition in electron-doped cuprates is driven by the weak-coupling d-wave BCS pairing in momentum space. In this case the critical
BCS temperature is given by Eq.~(\ref{3.62}):
\begin{equation}
\label{6.5}  
k_B \; T^{(e)}_{BCS}(\delta)  \;  \simeq \;   \frac{2  e^{\gamma}}{\pi}   \;  \Delta_2^{(e)}(0) \;  \exp{ [ -  \frac{1}{\lambda_2^{(e)}(\delta)}]}  
\end{equation}
with $\Delta_2^{(e)}(0)$ given by Eq.~(\ref{6.4}),
\begin{equation}
\label{6.6}  
\lambda_2^{(e)}  \; = \; \frac{m^*_e V_0}{\hbar^2} \; \int_{0}^{r_0^{(e)}(\delta)} dr \; r \left [ J_2(k_F^{(e)} r) \right ]^2 \; ,
\end{equation}
and
\begin{equation}
\label{6.7} 
 k_F^{(e)} \; \simeq \; \frac{\sqrt{ 2 \pi \delta}}{ a_0 } \; \;   . 
\end{equation}
The BCS critical temperature Eq.~(\ref{6.5}) as a function of the electron doping level is displayed  in Fig.~\ref{fig-13}. From the schematic phase diagram we infer
that the maximal critical temperature is strongly reduced with respect to the hole-doped cuprates. Moreover, the superconducting phase manifests itself in a rather
narrow interval of doping around $\delta \simeq 0.12$. These general aspects of our phase diagram seem to be in qualitative agreement with observations.
However, a thorough  understanding of the superconductivity in electron doped cuprates must await  more detailed experimental informations.
\section{Summary and Conclusions}
\label{7}
The driving principle of the approach presented in this paper has been that the high-temperature superconductivity can be understood within some framework along the
line of the microscopic theory of Bardeen, Cooper, and Schrieffer. The results presented in this paper summarize our attempts over the last years to construct a
comprehensive model that gives all aspects of the unusual behavior seen in the various regions of the phase diagram of the cuprates. It is obvious that we must rely heavily on some assumptions. First, we assumed that the physics of the high-temperature cuprates is deeply rooted in the copper-oxide planes. This allowed us to completely neglect the motion along the direction perpendicular to the $CuO_2$ planes. Our second main assumption has been that the single-band effective Hubbard model is sufficient to account for all the essential physics of the copper-oxide planes. Both assumptions are well accepted and superbly illustrated in the Anderson's 
book~\cite{Anderson:1997}. Accordingly, we have proposed an effective Hamiltonian that is able to govern the dynamics of the electrons or holes injected into the
undoped copper-oxide planes. We arrived at our effective Hamiltonian by using  arguments well-known in the literature on the motion of charge carriers in an
antiferromagnetic background. Notwithstanding, we were unable to offer a truly microscopic derivation of the effective Hamiltonian. So that our arguments to determine
the effective Hamiltonian, albeit suggestive, cannot be considered a first principle derivation. In spite of that, we showed that the effective Hamiltonian offers us a consistent picture of the high transition temperature cuprate superconductors. Due to the reduced dimensionality the two-body attractive potential admits real-space d-wave bound states. The binding energy of these bound states, which plays the role of the pseudogap, decreases with increasing doping until it vanishes at a certain critical doping  $\delta = \delta^*$. This allows us to reach the conclusion that the key features of the underdoped side of the phase diagram are controlled by very strong pairing that is phase-disordered by thermal fluctuations. We also were able to estimate the upper critical magnetic field whose strength and doping dependence are in good agreement with observations. We obtain a clear explanation for the so-called vortex region and for the puzzling discrete states bounded to the core of isolated vortices. We have discussed how the presence of the pseudogap is responsible for the formation of the quasielectron nodal Fermi liquid which, in turn, leads to the Fermi arcs observed in  angle-resolved photoemission studies and to the Fermi pockets in quantum oscillation experiments. We have found a natural explanation  for the peculiar dependence of the specific heat on the applied magnetic field observed in underdoped cuprates.  However, we would like to stress that we did not attempt
a complete discussion of the strange behavior of the cuprates in this region. In particular, as already pointed out in Sect.~\ref{4}, 
 we did not discuss the so-called nodal gap. In fact, we plan to present in a separate paper a detailed discussion on the origin and temperature dependence
 of the nodal gap, on the thermal conductivity, on the departure from the Wiedemann-Franz law and on the London penetration depth.  \\
In our model the overdoped region is realized for hole doping in excess of the critical doping $\delta^*$ where the pseudogap vanishes. In this region the conventional d-wave BCS framework account for many of the low-energy and low-temperature properties of the copper oxides, in agreement with several observations. Finally, we explain naturally why the superconducting critical temperature reaches its maximum in the optimal doped region. We pointed out that in this region the competition between the pseudogap and the d-wave BCS gap together  with the enhanced role of the phase fluctuations makes  questionable the usually adopted mean-field approximation. We have also briefly discussed the comparatively less-studied electron-doped cuprates. We suggest that the dome-shaped superconducting region in electron-doped copper-oxides can be described by the conventional weak-coupling d-wave pairing in momentum space. \\

In conclusion, the fact that a relatively simple model for the effective Hamiltonian allowed to recover several observational features of the high-temperature copper oxides,  suggests that the approach presented in this paper could be considered a significative step toward a microscopic explanation of the high-temperature superconductivity in cuprates.

\begin{thebibliography}{99}
%
\bibitem{Bednorz:1986}
J.~G. Bednorz and K.~A. M\"uller, Z. Phys. B {\bf 64} (1986) 189 
%
\bibitem{Damascelli:2003}
A. Damascelli, Z. Hussain  and Z.~X. Shen,  Rev. Mod. Phys. {\bf 75}  (2003) 473 
%
\bibitem{Deutscher:2005}
G. Deutscher,   Rev. Mod. Phys. {\bf 77} (2005)  109
%
\bibitem{Besov:2005}
D.~N. Besov and T. Timusk, ,  Rev. Mod. Phys. {\bf 77} (2005)  721
%
\bibitem{Lee:2006}
P.~A. Lee, N. Nagaosa, and X.~-G. Wen,   Rev. Mod. Phys. {\bf 78} (2006) 17
%
%
\bibitem{Fischer:2007}
\O. Fischer, H. Kugler, I. Maggio-Aprile  and C. Berthod,   Rev. Mod. Phys. {\bf 79} (2007) 353
%
%
\bibitem{Lee:2008}
P.~A. Lee ,  Rep. Prog. Phys.  {\bf 71} (2008)  012501
%
\bibitem{Hufner:2008}
S. H\"ufner, M.~A. Hossain, A. Damascelli and G.~A. Sarvatzky,   Rep. Prog. Phys.  {\bf 71} (2008) 062501
%
\bibitem{Sebastian:2012}
S. E. Sebastian,  N. Harrison  and G.~G. Lonzarich,   Rep. Prog. Phys.  {\bf 75} (2012) 102501
%
%
\bibitem{Anderson:1987}
P.~W. Anderson, Science    {\bf 235} (1987)  1196 
%
\bibitem{Anderson:1997}
P.~W. Anderson, {\it The Theory of Superconductivity in the High-$T_c$ Cuprates}, 
Princeton University Press, Princeton, New Jersey (1997)
%
%
\bibitem{Anderson:1959}
P.~W. Anderson, Phys. Rev.  {\bf 115} (1959)  2 
%
%
\bibitem{Bonn:2006}
D.~A. Bonn,  Nature Physics   {\bf 2}  (2006) 159
%
%
\bibitem{Niedermayer:1998}
Ch. Niedermayer  {\it et al.}, Phys. Rev. Lett.  {\bf 80} (1998)  3843
%
%
\bibitem{Cheong:1991}
S.~-W. Cheong   {\it et al.}, Phys. Rev. Lett.  {\bf 67} (1991) 1791
%
\bibitem{Mason:1992}
T.~E. Mason, G. Aeppli  and H.~A. Mook,    Phys. Rev. Lett.  {\bf 68} (1982)  1414
%
%
\bibitem{Lavrov:2009}
A.~N. Lavrov, L.~P. Kozeeva, M.~R. Trunin  and V.~N. Zverev,  Phys. Rev. B {\bf 79} (2009)  214523
%
%
\bibitem{Bardeen:1957}
J. Bardeen, L.~N. Cooper  and J.~R. Schrieffer,  Phys. Rev.  {\bf 106} (1957) 162;  Phys. Rev. {\bf 108} (1957) 1175
%
%
\bibitem{Armitage:2010}
N.~P. Armitage,  P. Fournier  and R.~L. Greene,   Rev. Mod. Phys. {\bf 82} (2010)  2421
%
%
\bibitem{Balachandran:1990}
A.~P. Balachandran, E. Ercolessi, and G. Morandi, {\it Hubbard Model and Anyon Superconductivity}, 
World Scientific Publishing Co. Inc. (1990) 
%
%
\bibitem{Huang:1987}
K. Huang and E. Manousakis,  Phys. Rev. B {\bf 36} (1987)  8302
%
%
\bibitem{Hirsch:1987}
J.~E. Hirsch,    Phys. Rev. Lett.  {\bf 59} (1987)  228
%
\bibitem{Trugman:1988}
S.~A. Trugman,  Phys. Rev. B {\bf 37} (1988) 1597
%
\bibitem{Carlson:2002}            
E.~W. Carlson, V.~J. Emery, S.~A. Kivelson  and D. Orgad,   {\it Concepts in High Temperature Superconductivity },
cond-mat/0206217 
%
%
\bibitem{Schrieffer:1988}
J.~R. Schrieffer,  X.~-G. Wen, and S.~-C. Zhang, Phys. Rev. Lett.  {\bf 60} (1988) 944;  Phys. Rev. B {\bf 39} (1989) 11663
%
%
\bibitem{Cooper:1956}
L.~N.  Cooper,   Phys. Rev.  {\bf 104} (1956) 1189
%
\bibitem{Schafroth:1955}
M.~R. Schafroth,   Phys. Rev.  {\bf 100} (1955) 463
%
\bibitem{Blatt:1955}
J.~M. Blatt and S.~T. Butler,   Phys. Rev.  {\bf 100} (1955) 476
%
\bibitem{Anderson:1966}
P.~W. Anderson, Rev. Mod. Phys.     {\bf 38}  (1966)  298
%
%
\bibitem{Legget:2008}
A.~J. Legget, {\it Quantum Liquids}, Oxford University Press, Oxford, UK (2008) 
%
%
\bibitem{Berezinskii:1971}
V.~L. Berezinskii, Sov. Phys. JETP {\bf 32} (1971) 493;   Sov. Phys. JETP {\bf 34} (1972)  610
%
\bibitem{Kosterlitz:1973}
J.~M. Kosterlitz and D.~J. Thouless, J. Phys. C  {\bf 6} (1973)  1181
%
%
\bibitem{Vignolle:2008}
B. Vignolle   {\it et al.},  Nature   {\bf 455} (2008)  952
%
\bibitem{Anderson:1961}
P.~W. Anderson and P. Morel,  Phys. Rev.    {\bf 123} (1961) 1911
%
%
\bibitem{Won:1994}
H. Won   and K. Maki,  Phys. Rev.  B  {\bf 49} (1994) 1397
%
\bibitem{Timusk:1999}
T. Timusk   and B. Statt,  Rep. Prog. Phys.   {\bf 62} (1999)  61
%
\bibitem{Tallon:2001}
J.~L. Tallon   and J.~W. Loran,  Physica  C  {\bf 349} (2001) 53 
%
%
\bibitem{Chatterjee:2010}
U. Chatterjee   {\it et al.},  Nature Physics   {\bf 6} (2010)  99
%
%
\bibitem{Tanaka:2006}
K. Tanaka  {\it et al.},  Science   {\bf 314} (2006)  1910
%
%
\bibitem{He:2009}
R.-H. He   {\it et al.},  Nature Physics   {\bf 5} (2009)  119 
%
\bibitem{Pushp:2009}
A. Pushp   {\it et al.},  Science   {\bf 324} (2009)  1689
%
%
\bibitem{Hashimoto:2010}
M. Hashimoto   {\it et al.},  Nature Physics   {\bf 6} (2010)  414 
%
%
\bibitem{Reber:2012}
T. J. Reber   {\it et al.},  Nature Physics   {\bf 8} (2012)  606 
%
%
\bibitem{Sakai:2012}
S. Sakai {\it et al.},  {\it Exploring the Dark Side of Cuprate Superconductors: s-wave Symmetry of the
Pseudogap},  arXiv:1207.5070 
%
%
\bibitem{Yoshida:2012}
T. Yoshida  {\it et al.},  {\it Coexisting pseudo-gap and superconducting gap in the high-T$_c$ superconductor 
 La$_{2-x}$Sr$_{x}$CuO$_{4}$}, arXiv:1208.2903 
%
%
\bibitem{Hashimoto:2012}
M. Hashimoto   {\it et al.},  Phys. Rev. B {\bf 86}  (2012)  094504 
%
%
\bibitem{Kugler:2001}
M. Kugler  {\it et al.},  Phys. Rev. Lett.  {\bf 86} (2001) 4911
%
%
\bibitem{Hetel:2007}
I. Hetel, T.~R. Lemberger  and M. Randeira,  Nature Physics   {\bf 3} (2007)  700
%
\bibitem{Corson:1999}
J. Corson  {\it et al.},   Nature   {\bf 398} (1999) 221
%
%
\bibitem{Emery:1995}
V.~J. Emery and S.~A. Kilvelson,   Nature    {\bf 374} (1985)  434
%
%
\bibitem{Tinkham:1996}
M. Tinkham,  { \it Introduction to Superconductivity}, Second Edition, Mc Graw-Hill, Inc., New York (1996)
%
%
\bibitem{Wang:2006}
Y.~ Wang, L.~Li  and N.~P.~Ong, Phys. Rev. B {\bf 73} (2006)  024510
%
%
\bibitem{Li:2007}
L.~Li  {\it et al.},  Nature Physics {\bf 3} (2007)  311
%
%
\bibitem{Chang:2012}
J.~Chang  {\it et al.}, Nature Physics {\bf 8} (2012)  751
%
%
\bibitem{Xu:2000}
Z.~A. Xu  {\it et al.},  Nature {\bf 406} (2000)  486 
%
\bibitem{Capan:2002}
C. Capan   {\it et al.},  Phys. Rev. Lett.  {\bf 88} (2002)  056601
%
%
\bibitem{Li:2007b}
L.~Li,  Y. Wang, J.~G. Checkelsky  and M.~J. Naughton, Physica C  {\bf 460-462} (2007)  48
%
%
\bibitem{Caroli:1964}
C. Caroli, P.~G. de Gennes  and J. Matricon,  Phys.  Lett.  {\bf 9} (1964)  307
%
%
\bibitem{Doiron:2007}
N. Doiron-Leyrand  {\it et al.},  Nature {\bf 447} (2007)  565
%
%
\bibitem{Yelland:2008}
E.~A. Yelland  {\it et al.},  Phys. Rev. Lett.  {\bf 100} (2008)  047003
%
%
\bibitem{Bangura:2008}
A.~F. Bangura   {\it et al.},  Phys. Rev. Lett.  {\bf 100} (2008)  047004
%
\bibitem{Jaudet:2009}
C. Jaudet   {\it et al.},  Physica B    {\bf 404} (2009)  354
%
\bibitem{Bangura:2010}
A.~F. Bangura   {\it et al.},  Phys. Rev. B  {\bf 82} (2010)  140501
%
%
\bibitem{Abrikosov:1972}
A.~A. Abrikosov,  { \it Introduction to Theory of Normal Metal}, Academic Press, New York and London  (1972) 
%
%
\bibitem{Shoenberg:1984}
D. Shoenberg,  { \it Magnetic Oscillations in Metals}, Cambridge University Press, Cambridge  (1984)
%
\bibitem{LeBoeuf:2007}
D. Le Boeuf  {\it et al.},  Nature {\bf 450} (2007)  533 
%
%
\bibitem{Sebastian:2011}            
S. Sebastian  {\it et al.},  	Nature Communications {\bf 2} (2011) 471 
%
\bibitem{Laliberte:2011}            
F. Laliberte  {\it et al.},  Nature Communications {\bf 2} (2011) 432 
%
%
\bibitem{Riggs:2011}
S.~C. Riggs  {\it et al.},  Nature Physics  {\bf 7} (2011)  332 
%
%
\bibitem{Ashcroft:1976}
N.~W. Ashcroft and N.~D. Mermin,  {\it Solid State Physics},  Harcourt College Publishers  (1976) 
%
%
\bibitem{Volovik:1993}
G.~E. Volovik, Sov. Phys. JETP {\bf 58} (1993) 469   
%
\bibitem{Wade:1994}            
J.~M. Wade  {\it et al.},  J. Supercond.  {\bf 7} (1994)  261  
%
\bibitem{Wang:2007}  
Y.~ Wang   {\it et al.},  Phys. Rev. B {\bf 76} (2007)  064512
%
\bibitem{Guyard:2008}  
W. Guyard   {\it et al.},  Phys. Rev. B {\bf 77} (2008)  024524
%
\bibitem{Loram:1997}
J.~W. Loram, K.~A. Mirza, J.~R. Cooper  and J.~L. Tallon,  Physica C    {\bf 282-287} (1997)   1405
%
%
\bibitem{Loram:2001}
J.~W. Loram, J. Luo, J.~R. Cooper  and J.~L. Tallon,  J. Phys. Chem. Sol.     {\bf 62} (2001)  59
%
%
\bibitem{Daou:2009}
R. Daou  {\it et al.},  Nature Physics {\bf 5} (2009)  31 
%
%
\bibitem{Daou:2009b}
R. Daou  {\it et al.},  Phys. Rev. B  {\bf 79} (2009)  180505
%
\end{thebibliography}
\end{document}